# High repetition rate ultrafast electron diffraction with direct electron detection

F. Rodriguez Diaz[1], M. Mero[1], K. Amini[1,*]

[1]Max-Born-Institut, Max-Born-Straße 2A, 12489 Berlin, Germany

*corresponding author e-mail: amini@mbi-berlin.de

## ABSTRACT

Ultrafast electron diffraction (UED) instruments typically operate at kHz or lower repetition rates and rely on indirect detection of electrons. However, these experiments encounter limitations because they are required to use electron beams containing a relatively large number of electrons (>>100 electrons/pulse), leading to severe space-charge effects. Consequently, electron pulses with long durations and large transverse diameters are used to interrogate the sample. Here, we introduce a novel UED instrument operating at a high repetition rate and employing direct electron detection. We operate significantly below the severe space-charge regime by using electron beams containing 1 to 140 electrons per pulse at 30-kHz. We demonstrate the ability to detect time-resolved signals from thin film solid samples with a difference contrast signal, $\Delta I/I_0$, and an instrument response function as low as $10^{-5}$ and 184-fs (FWHM), respectively, without temporal compression. Overall, our findings underscore the importance of increasing the repetition rate of UED experiments and adopting a direct electron detection scheme, which will be particularly impactful for gas-phase UED. Our newly developed scheme enables more efficient and sensitive investigations of ultrafast dynamics in photoexcited samples using ultrashort electron beams.

## I. INTRODUCTION

Ultrafast electron diffraction (UED)[1–25] is a powerful technique that tracks changes in the position of atoms within a material in real-time with picometre and femtosecond spatio-temporal resolution. UED is often employed in pump-probe configuration where an optical ("pump") pulse photoexcites a sample away from its ground-state structure and another, time-delayed, electron ("probe") pulse measures diffraction patterns of the excited sample. Pulses of electrons with high kinetic energies (in the keV or MeV) are easily attainable, providing an electron probe pulse with a (sub-)picometre de Broglie wavelength[26–28]. The Fourier transform of the resulting diffraction pattern yield structural information with a spatial resolution of less than 10-pm[18,21,23,28]. Measurement of diffraction patterns at various time delays between the two pulses allows the retrieval of a real-space "movie" of structural changes in the excited sample during a photochemical reaction.

In the early 1980s, Mourou, Williamson and Li introduced the first picosecond variant of electron diffraction in transmission mode.[1,2] The development of femtochemistry by Zewail[29,30] enabled the use of femtosecond optical pulses to generate electron pulses[31,3,7]. Despite this advancement, space-charge dispersion persisted in limiting the electron pulse duration to the picosecond regime.[31,3,7] Following the ground-breaking contributions of Mourou[1,2] and Zewail[31,3,7], extensive efforts have been dedicated to advancing the UED technique by numerous research groups and facilities. Figure 1 illustrates the impact of space-charge dispersion and the performance of different UED set-ups implementing diverse strategies. To reach a high spatio-temporal resolution, it is particularly important to generate an ultrashort electron pulse with the lowest pulse duration and transverse electron beam diameter. The electron pulse duration often dictates the total temporal resolution of UED measurements, referred to as the instrument response function (IRF). Reducing the transverse diameter of electron beams on-target minimizes the optical pump pulse diameter and, consequently, lowers the average power requirements of laser systems, which becomes particularly important when operating at high repetition rates. Furthermore, maintaining a high average beam current at the sample position is equally crucial to ensure the measurement of scattering signals with a high signal-to-noise (SNR) ratio. This is particularly crucial in gas-phase UED studies, where electron scattering signals are approximately 100-1000 times weaker than those measured with solid thin films. Achieving the necessary high average beam currents in gas-phase UED is feasible by operating at high repetition rates. While cumulative heating effects may not be significant when studying simple metallic films, they become problematic when investigating more complex solid-state films that require a low base temperature[32]. Our work aims to demonstrate the technical performance of our high repetition rate UED instrument using simple thin metallic films, in preparation for its use for future gas-phase UED studies, which are well-suited for operation at high repetition rates.

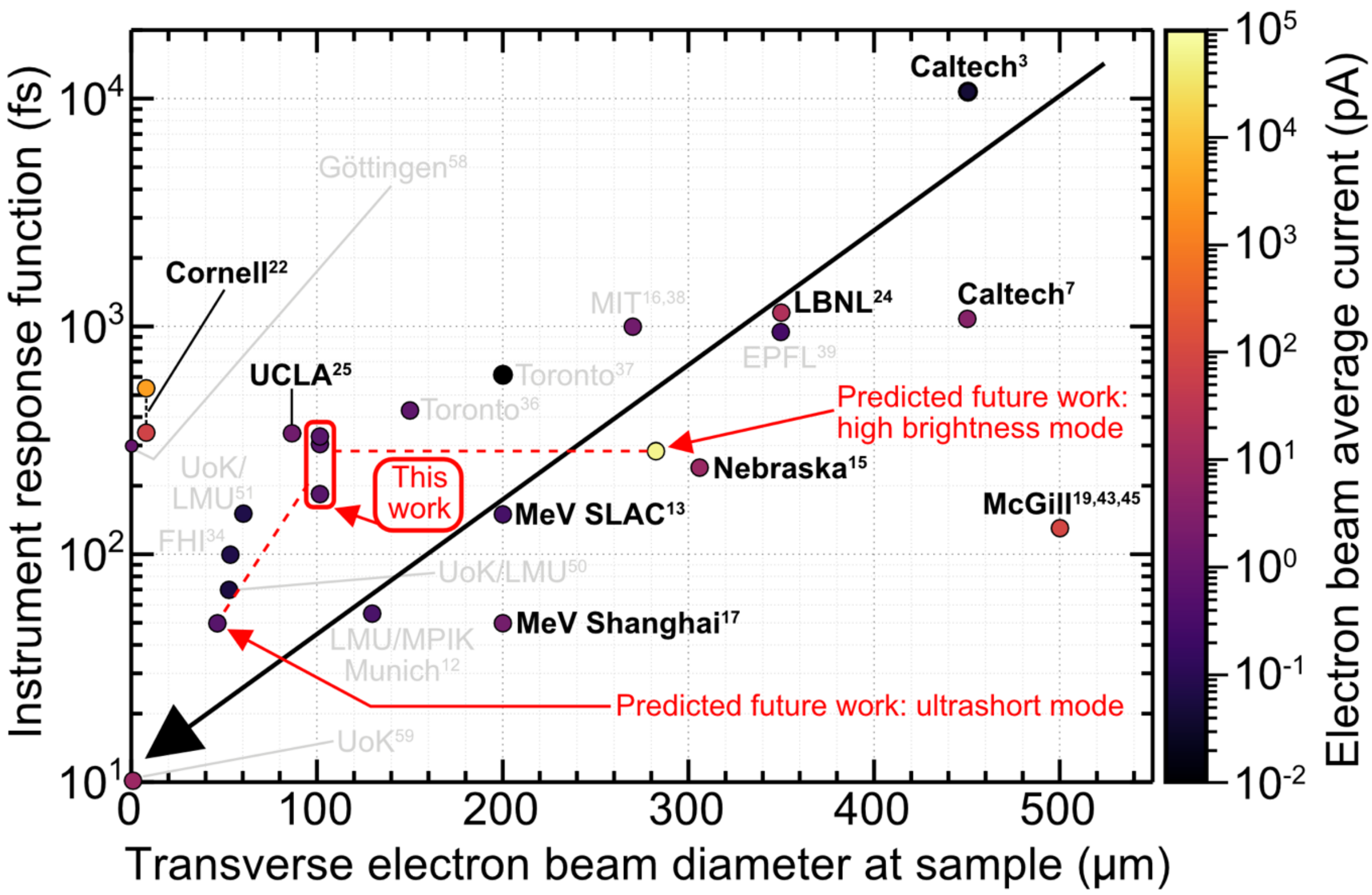

**FIG. 1.** Evolution of instrument response function and transverse electron beam diameter at the sample position as a function of the average current of the electron beam. All values are given as full width at half maximum (FWHM) values. The diagonal black arrow indicates the evolution of UED towards shorter and smaller electron pulses. The brightness of the electron beam is depicted by the $z$-scale colourmap representing the electron beam average current. Set-ups that are generally incompatible for gas-phase UED measurements are shown in grey text. Our work is highlighted by red text, arrows and rectangles. The anticipated future capabilities of our set-up are depicted by dashed lines, indicating the expected range of parameters.

Realizing such optimal electron beam characteristics poses significant challenges owing to various constraints. These include limitations imposed by low repetition rates and severe space-charge effects, as well as issues associated with low electron scattering signals, non-optimal electron detection methodologies, and constraints related to electron beam flux and brightness. The relatively low repetition rate of laser systems driving UED set-ups necessitates the use of electron pulses with relatively high bunch charges (>>16-aC), leading to an increase in pulse duration from ~150-fs to more than 1-ps in one metre propagation due to severe space-charge dispersion for keV electrons. Siwick, Miller and coworkers were the first to demonstrate sub-picosecond UED measurements.[33] In these solid-state keV UED studies, space-charge effects were overcome by minimizing the sample-to-photocathode distance[33], often to less than 5-cm, and improving the initial electron beam characteristics.[34,35] This optimization yielded an IRF as short as 100 fs and transverse beam diameters as small as less than 100-µm.[34] However, instruments employing short sample-photocathode distances[16,34,36–39] are generally incompatible with gas-phase UED experiments due to the higher operating pressures and the associated risk of voltage breakdown in the electron accelerator. So far, the generation of ultrashort electron pulses with durations of 100 fs or less over longer electron propagation distances has become feasible through several advancements. These include the utilization of relativistic MeV beams[9,13,14,17,20,21,40,41] or the implementation of temporal compression schemes employing radiofrequency (RF) fields generated by a microwave cavity[42,10,43,44,19,45] or optical terahertz (THz) fields[12,46–48]. Space-charge effects in 3-MeV electron beams are three orders-of-magnitude weaker than compared to <100-keV electrons at comparable bunch charge.[13] This consequently enabled MeV-UED experiments with sub-150-fs time resolution.[9,13,14,17,20,21,40,41] Notably, MeV-UED measurements with an IRF of sub-50-fs are now possible.[17] However, access to accelerator-based MeV electron beams is confined to large-scale facilities, and achieving sub-10-fs time resolution with MeV set-ups necessitates precise synchronization of the RF accelerating field with the pump laser pulse[20,25,43,49]. In keV UED instruments with a bunch charge much greater than 16-aC, the highest temporal resolution (FWHM) reported so far is approximately 100-fs in solid-state studies[19,50,51] and 240-fs in gas-phase work[15]. Furthermore, notable achievements include temporal compression of single-electron pulses to 28-fs (FWHM)[10], the bunching of a 50-electron pulse into a train of 800-as single-electron pulses[12], and the use of the optical gating approach for attosecond electron diffraction[52–54].

Additionally, ultrafast variants of transmission electron microscopes[53–59] (UTEMs) have demonstrated electron pulses ranging from hundreds of femtoseconds[58] to attoseconds[60,59] duration, with nanometre-scale transverse focussed beam diameters[24,58,61] reaching down to 9 Å.[58] Performing gas-phase measurements using UTEMs is technically challenging. Therefore, in this work, we will focus solely on UED set-ups capable of performing gas-phase UED measurements. It is important to note that in electron beam physics, the 6D bunch brightness[57] is generally a better figure of merit than the electron flux. This is particularly crucial for solid thin film samples, which suffer from thermal effects at relatively high repetition rates and require small (few µm or less) electron beam diameters, along with tailored sample thin films using nm-scale gold masks[62]. However, for gas-phase measurements, the smallest transverse beam size does not necessarily need to be on the few-µm or smaller scale. This is because the length of a typical gas jet is at least 100-200 µm[13,44,63], and even extend to the millimetre scale in the case of MeV beams using flow cells[63]. To achieve suitable a signal-to-noise ratio in gas-phase UED, a minimum transverse beam diameter of at least approximately 50-100 µm is ideally desired. Such a transverse beam diameter will still be sufficient to alleviate the average power (≥200-W) requirements for performing UED measurements with a low bunch charge (<16-aC)[12,50,51] at higher repetition rates (≥100-kHz) than those currently employed.

Electron detection with a minimal noise contribution to the SNR ratio is another crucial factor. High-energy electron beams are often detected using a phosphor scintillator screen, which emits a few hundred photons for each detected electron.[64] The emitted photons are subsequently measured by either a fiber- or lens-coupled photon-sensitive imaging sensor. These sensors, which include charge-coupled device (CCD)[65], electron-multiplying CCD (EMCCD)[13], or complementary metal oxide semiconductor (CMOS)[64] technology, are susceptible to saturation effects from bright signals. For example, saturation from the unscattered beam is typically mitigated by employing a beam blocker positioned in-front of the phosphor screen. However, this approach results in the loss of valuable information on fluctuations in the electron beam's intensity and pointing, which are crucial for correcting these fluctuations. Furthermore, these charge-integrating analog detectors also suffer from relatively larger gain, integration and readout noise, limiting the SNR ratio that can be achieved.[64] Additionally, these indirect electron detection schemes often exhibit a limited dynamic range, particularly at high momentum transfers. It is worth noting that microchannel plates (MCPs), positioned in-front of a phosphor scintillator screen[65], have also been used, however, their detection efficiency of high-energy electron beams (<15%) is not ideal.[66]

In this work, we employ a novel, high repetition rate UED (HiRepUED) instrument operating at 30-kHz, utilizing direct electron detection. Our approach involves the use of temporally uncompressed electron pulses containing between 55 – 140 electrons, corresponding to a bunch charge of 8 – 24 aC. This bunch charge is significantly lower than that employed in existing (sub-)kHz UED set-ups (2-20 fC), leading to significantly weaker space-charge forces and lower emittances in keV beams, allowing us to operate in the low-to-moderate space-charge regime. We find that temporally uncompressed electron pulses as short as 174-fs are achievable, resulting in an IRF of ~184-fs. This IRF is better than that reported for the keV UED instrument utilized by Centurion and coworkers (240-fs) but using a compressed electron scheme[15]. Furthermore, we demonstrate the measurement of time-resolved signals with improved SNR ratio using direct electron detection as compared to indirect detection schemes. As a future outlook, we also discuss the future potential of our instrument with RF temporal compression, and give a brief discussion of its anticipated performance when operating in the ultrashort and ultrabright modes (see dashed lines in Fig. 1).

This paper is structured as follows. Details of the experimental set-up and simulations are given in Sections II and III, respectively. Results obtained from measurements and simulations are discussed in Section IV, and a summary conclusion is given in Section V.

## II. EXPERIMENTAL

### A. Optical set-up

Figure 2 shows a schematic of the optical set-up employed in this study. Here, the amplified output of a 30-kHz, 200-μJ laser system (Light Conversion PHAROS, 1025-nm, 6-W, 260-fs) was split into two beams using a half waveplate (HWP) and a thin film polarizer (TFP) positioned at Brewster's angle (see Fig. 2a). The $p$-polarized transmitted pulse was used as a supercontinuum seed pulse for white light supercontinuum generation (WLG) in the home-built optical parametric chirped-pulsed amplification (OPCPA) set-up. The $s$-polarized reflection was frequency converted to 512.5-nm (220-fs FWHM, 100-μJ) and was used as the pump beam of two non-collinear optical parametric amplification (NOPA) stages. The footprint of the OPCPA was 60 × 45 $cm^2$. The signal output of the OPCPA ($\lambda_c$ = 800-nm, ~17-μJ, ~190-fs) was temporally compressed to 49 fs using prism compression (see Fig. 2b) close to its Fourier transform limit (TL) of 47-fs (19.5-nm FWHM bandwidth, see Fig. 2b). The compressed near-infrared (NIR) 800-nm pulse was split into two beams using an 80:20 beam splitter (see Fig. 2c) where one beam was used for electron generation while the second beam was used for sample photoexcitation.

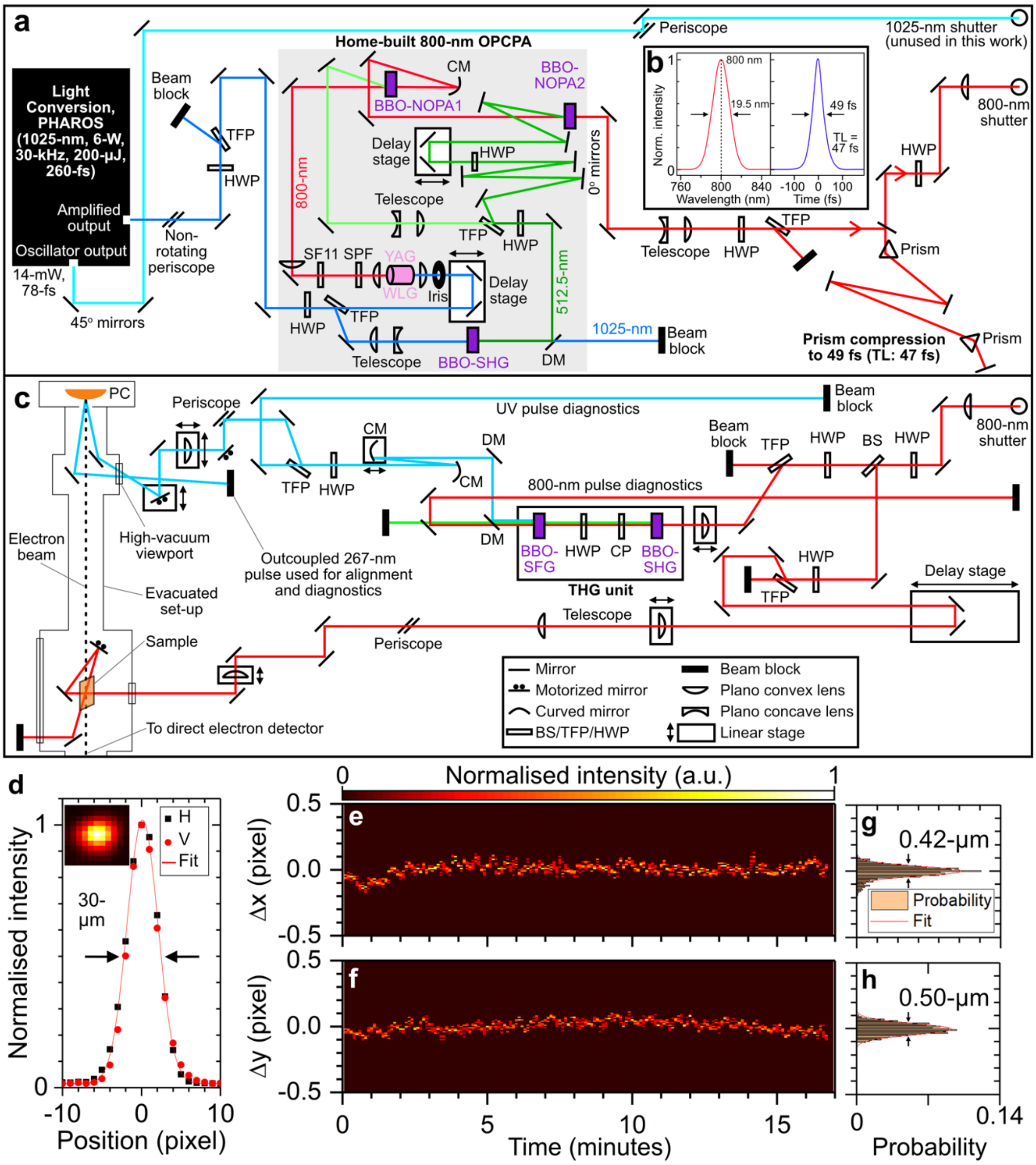

**FIG. 2. (a)** Schematic of a 30-kHz, home-built optical parametric chirped pulse amplification (OPCPA) system that employs a 6-W PHAROS laser system as the supercontinuum-seed and pump laser, and subsequent prism compression. **(b)** Wavelength and temporal profile of OPCPA output reconstructed from frequency-resolved optical gating (FROG) measurements. The central wavelength, $\lambda_c$, and transform limit (TL) of the pulse are indicated. **(c)** Schematic of optical set-up for generating electron probe pulses using ultraviolet (UV) light at the photocathode (PC), and sample excitation using 800-nm optical pump pulses. **(d)** UV beam profile. Cuts along the horizontal (H) and vertical (V) axes are shown with a Gaussian fit. Inset: UV beam detector image. **(e-f)** UV beam pointing measurement. Changes in the (b) horizontal ($\Delta x$) and (c) vertical ($\Delta y$) position of the UV beam are shown. **(g-h)** Histogram distribution of the corresponding data shown in panel (b-c). Gaussian fits were applied to the histogram distributions.

In the electron generation (probe) beam, the NIR pulse (3-μJ) was frequency converted to 267-nm (20-nJ, 90-fs, TL: 42-fs) using a compact third harmonic generation unit (Eksma Optics, FemtoKit) composed of two BBO crystals, a HWP and a calcite plate (CP). Two harmonic separating dichroic mirrors (DMs) separated the 267-nm ultraviolet (UV) pulse from the residual 400-nm and 800-nm pulses. The UV pulse was subsequently beam expanded using two curved mirrors (CMs) by a magnification factor of two, producing a collimated UV beam (~5-mm FWHM diameter). The UV pulse was then power attenuated using a HWP and TFP positioned at Brewster's angle. The $p$-polarized transmitted pulse was used for UV pulse diagnostics and alignment purposes. The $s$-polarized reflection was directed to the UED instrument, where it was focussed to a 30-μm FWHM diameter spot on the photocathode (see beam profile in Fig. 2d). The ultraviolet (UV) pulse duration at the photocathode was estimated to be 90-fs based on the group dispersion delay (GDD; ~1200-fs$^2$) expected from the UV optics employed in our optical set-up (see Fig. 2b). The pointing stability of the UV pulse at the focus was measured to be ~0.50-μm using a UV-sensitive beam profiling camera (PCO.edge 4.2 bi UV) over a ~15-minute period, as shown in Fig. 2e-h. Additionally, an intensity jitter of 1% (FWHM) was observed during this period. As the UV pulse is generated through non-linear conversion using an 800-nm pulse, the above-mentioned jitter values of the UV pulse can be taken as the maximum corresponding values for the 800-nm pulse. Notably, our optical setup does not employ a beam pointing stabilization system.

In the sample photoexcitation (pump) beam, the NIR pulse energy (12-μJ) was attenuated using a HWP and TFP at Brewster's angle, and subsequently reflected by a retroreflector mounted on a linear delay stage (Physik Instrumente M-531.2S1, 30-cm travel range). The NIR pulse was beam expanded and subsequently focussed to a ~180-μm diameter spot at the sample position using a plano-convex uncoated lens ($f$ = +600-mm). The relative angle of the optical pump pulse to the electron axis was ~20°. The mirror after the NIR lens and the last mirror before the sample are piezo-mounted to obtain the optimal spatial overlap with the electron beam at the sample position.

## B. Experimental set-up

Figure 3 shows a schematic of the HiRep-UED instrument employed in this study. The UV pulse was focussed to a 30-μm FWHM diameter spot (see Fig. 2d) on a high purity oxygen-free copper photocathode (>99.99%)[67]. Since the photocathode was held at a high voltage of 95-keV with a high voltage power supply (Matsusada AU-100N1.5-L), the effective work function of copper was reduced due to the Schottky effect (~4.3-eV)[67] such that the photon energy of the UV pulse (4.61-eV) was sufficiently high to generate electrons via photoemission. A grounded anode flange positioned ~11-mm away from the photocathode generated a constant direct current (DC) field with a field strength of $E_{DC}$<10 MV/m, sufficient to accelerate the electron beam to 95-keV. The electron beam passed through a 10-mm aperture and was transversely collimated and focussed by two solenoid magnetic lenses (MLs) labelled ML1 and ML2, respectively. The collimating solenoid magnetic lens ML1 (Doctor X Works) was placed directly in-front of the anode flange located 40 mm away from the photocathode. ML1 consisted of rectangular (2.5 x 1.5 mm) copper wire wound into a coil with 351 turns, an inner and outer diameter of 42.5-mm and 68.5-mm, respectively, and a length of 60-mm. The condensing solenoid magnetic lens ML2 (Doctor X Works) was placed ~670-mm away from the photocathode. ML2 comprised of circular (1-mm diameter) copper wire wound with 650 turns forming a 60-mm length coil with an inner and outer radius of 20-mm and 37.5-mm, respectively. The centre hole is 35-mm in diameter. A maximum on-axis magnetic field flux density of 40-mT (17.5-mT) at 12 A (1.75 A) for ML1 (ML2) can be generated. The focussed electron beam scattered against thin-film solid-state samples held by a 0.5-mm thick copper block capable of holding up to nine samples with 3-mm diameter. The sample holder was mounted onto a manual four-axis ($x$, $y$, $z$, $\theta_{yaw}$) ultrahigh vacuum mechanical translator stage (VAb Vakuum-Anlagenbau). A radiofrequency (RF) microwave cavity (Doctor X Works) for temporal electron compression was installed together with an active synchronization system that corrects the RF-laser timing jitter based on Ref.[43]. However, both are unused in this work as it is beyond the scope of the current work and is subject of a separate publication.

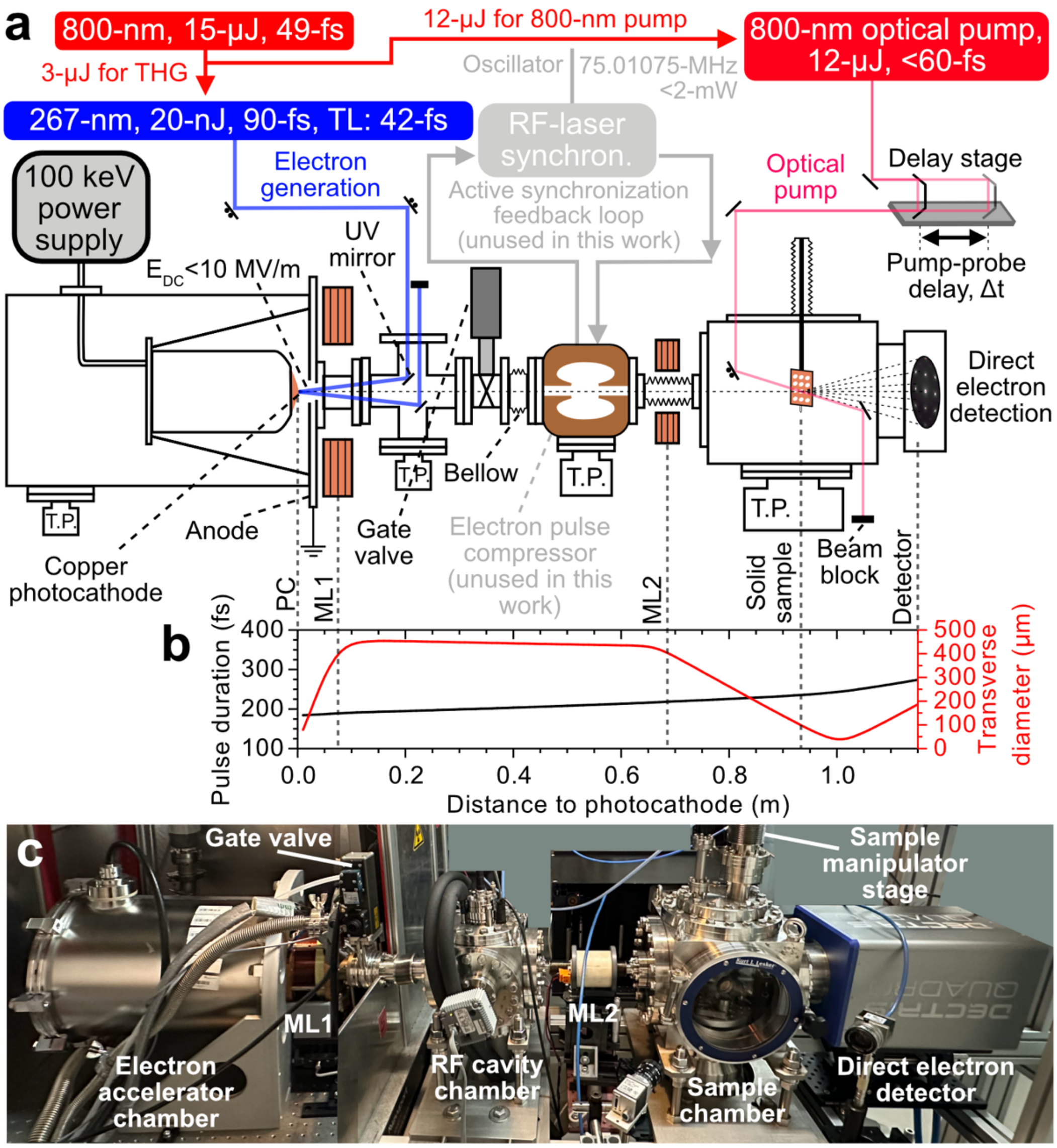

**FIG. 3. (a)** Schematic of the 30 kHz, 95 keV HiRep-UED instrument, with each component labelled. See main text for more details. **(b)** General Particle Tracer (GPT) simulation of the electron pulse duration (blue distribution) and transverse diameter (red distribution) with the instrument operating in the temporally uncompressed mode using an electron beam containing 55 electrons. All values are given in FWHM. **(c)** Image of the instrument in the laboratory.

The scattered and unscattered electrons are measured with a direct electron detector (DECTRIS QUADRO)[68,69] composed of 512 × 512 pixels with 75 μm × 75 μm size, giving a total active area of 38.4 mm × 38.4 mm. Each pixel was composed of a 450-μm thin silicon wafer. This combination ensures a detection efficiency of more than 90% for 95-keV electrons (*i.e.*, single-electron detection). Each pixel can detect up to four electrons per pulse (see Section IV.B), while possessing a 32-bit dynamic detection range (sum of two 16-bit electron counters), capable of counting up to 4.2 x $10^9$ hits in a given exposure time. The detector's maximum count rate was determined as $10^7$ electrons/second/pixel under continuous illumination.[68] Loss of sensitivity was observed with the detection of an unfocussed electron beam containing more than $10^4$ electrons at 30-kHz, with the beam covering the full sensor dimensions. In addition to the electron detector, a home-built Faraday cup (not shown in Fig. 3; 2-mm open aperture diameter, 20-mm length) positioned in-front of the detector was used to measure the average current of the electron beam containing more than $10^4$ electrons. Practical acquisition times for time-resolved measurements at 30-kHz spanned from between 20 to 90 minutes for data shown in this work. Typical instrument parameters of the HiRep-UED set-up are summarized in Table I. Further details of the HiRep-UED instrument are provided in Section B of the supplementary material.

**Table I.** Typical machine parameters for the HiRep-UED set-up in this work.

| **Parameters** | **Values** |
|---|---|
| Repetition rate | 30-kHz |
| Vacuum in the following chambers | |
| Electron accelerator | <7E-7 mbar |
| RF cavity | <3E-8 mbar |
| Sample chamber | <3E-8 mbar |
| Electron beam kinetic energy | 95-keV |
| Electron beam charge | 0.16-aC to 22-aC (1 - 140 electrons/pulse); overall range is 1.6-aC to 1.6-pC (i.e., 1 to $10^6$ electrons/pulse) |
| Electron flux | 1.5 x $10^6$ to 4.2 × $10^6$ electrons/second, (maximum of 3.0 × $10^{10}$ electrons/second) |
| At the photocathode position | |
| UV pulse size (FWHM) | 30-μm |
| UV pulse duration (FWHM) | 90-fs (TL: 42-fs) |
| UV pulse energy | <20-pJ (overall range of 1-pJ to 20-nJ) |
| At the sample position | |
| Electron beam charge (after passing through aperture) | 1.4-aC to 6.8-aC (9 to 42 electrons/pulse) |
| Electron bunch length (FWHM) | 174-fs to 322-fs (uncompressed), (<50-fs compressed predicted by GPT) |
| Electron beam size (FWHM) | ~100-μm (200-μm) |
| Electron beam transverse pointing jitter (FWHM) | ≤28-μm |
| Transverse coherence length (RMS) | 3.8-nm |
| Transverse emittance (RMS) | 3.8 nm·rad |
| Longitudinal emittance (RMS) | 103 fm·rad to 177 fm·rad |
| Pump laser spot size (FWHM) | ~180-μm |
| Pump laser duration (FWHM) | <60-fs (TL: 47-fs) |
| At the detector position | |
| Electron beam size (FWHM) | ~450-μm (280-μm) |
| Reciprocal-space resolution | 0.063-Å$^{-1}$ |
| Spatial resolution | 3.8-pm |

## III. SIMULATIONS

### A. General Particle Tracer simulations

General Particle Tracer (GPT)[70–72] simulations of the UED instrument were performed using an electron beam with a kinetic energy of 95-keV, and an excess thermal energy of 0.5-eV. The electron beam was generated by a UV pulse with a FWHM diameter and duration of 30-μm and 90-fs, respectively, without the use of the RF cavity (*i.e.*, temporally uncompressed mode). For details about the influence of the UV pulse duration on the temporally uncompressed electron pulse, the reader is referred to Section C and Figure S2 of the supplementary material. For comparison, simulations were also performed with RF-compressed electron beams (*i.e.*, temporally compressed mode), using a UV pulse duration of 60-fs. Field maps of the electron accelerator and solenoid magnetic lenses, generated from finite element method (FEM) simulations, were used. For each simulation, we used 10,000 macroparticles with the mesh method.

## IV. RESULTS AND DISCUSSIONS

### A. Electron beam characterization

#### *1. Electron flux*

Our UED set-up is capable of generating an electron pulse containing between one and >$10^6$ electrons at 30-kHz with a UV pulse energy between ~1-pJ to 20-nJ, as shown in Fig. 4a. Operating in high-brightness mode, our setup achieves an unprecedented electron flux exceeding $10^{10}$ electrons/second, surpassing existing keV-scale UED systems by more than an order of magnitude[15,19,22,25,44]. Compared to the brightest MeV-scale UED set-up,[13] our system has a 100-fold higher electron flux. Considering the factor of four higher elastic electron scattering cross-section for 95-keV electrons compared to 3-MeV electrons, calculated with ELSEPA,[73] the maximum electron scattering signal from our set-up is approximately 400 times greater than the current state-of-the-art in MeV-UED[13]. Even in anticipation of upgrades to MeV-UED facilities aiming for a 1-kHz repetition rate, the difference in signal level between our setup and MeV-UED remains appreciably high (factor of ~150).

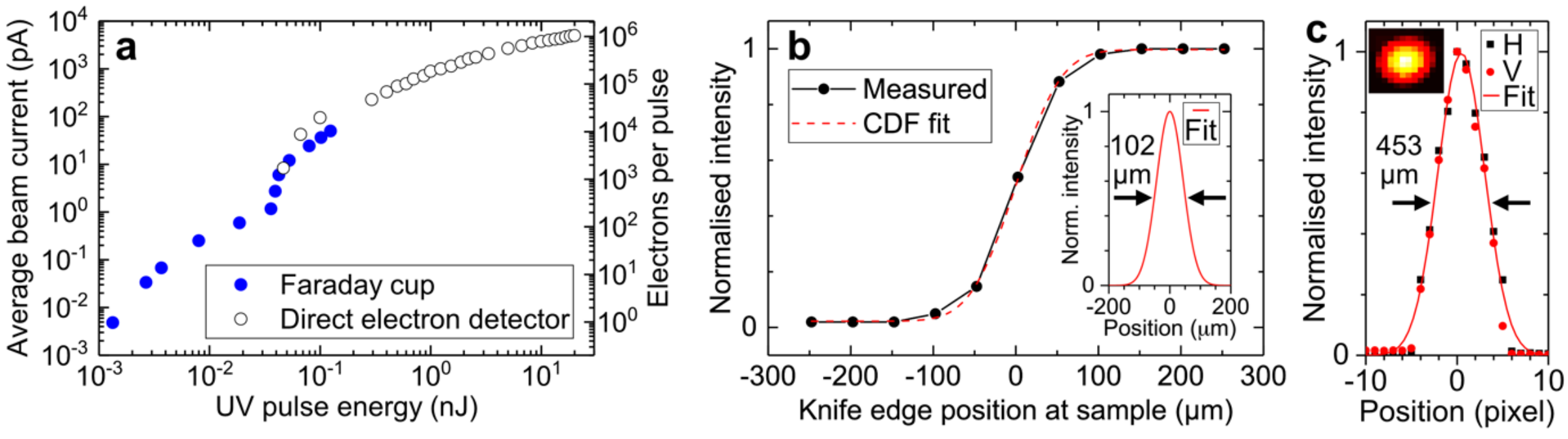

**FIG. 4. (a)** Average beam current and electrons per pulse as a function of UV pulse energy measured with a Faraday cup and direct electron detector, respectively. **(b)** Knife edge scan of electron beam at the sample position. A Gaussian cumulative distribution function (CDF) was fitted to the measured data. Inset: retrieved Gaussian distribution from CDF fit in panel (b). The FWHM electron beam diameter is indicated by black arrows. **(c)** Electron beam profile at the detector position obtained from the same data as in panel (b). Inset: electron detector image.

*2. Electron transverse profile at sample and detector*

We performed measurements to determine the full-width at half-maximum (FWHM) diameter of an electron beam containing 100 electrons. Using specific combinations of ML1 and ML2 currents, we adjusted the transverse focus of the beam to be positioned either close to the sample or the detector. Knife-edge scans were employed to measure the electron beam size at the sample position. The electron beam size at the detector was measured directly on the direct electron detector. In our typical operating configuration, with the transverse focus at the sample, we measured beam diameters of 102-μm and 453-μm at the sample and the detector, respectively (see Fig. 4b-c). With a different combination of ML1 and ML2 currents, we were able to reduce the electron beam diameter to approximately 280-μm at the detector, albeit with a larger 200-μm diameter at the sample. Figure 3b shows corresponding General Particle Tracer (GPT) simulations of the transverse diameter of an electron pulse containing 55 electrons in our UED instrument (*i.e.*, 8.8-aC; see red line). The initially divergent electron beam is collimated by ML1 to a diameter of 450-μm (FWHM) at the position of the RF cavity (~0.5-m from photocathode). The collimated beam is subsequently focussed by ML2 to 100-μm (FWHM) close to the sample position (~0.95-m from photocathode). A good agreement between the measured (102-μm) and simulated (100-μm) transverse beam diameter at the sample position is observed. The GPT simulations further predict an RMS transverse and longitudinal emittance at the sample of 5.6 nm·rad and 103 fm·rad (10.5 nm·rad and 11.4 pm·rad), respectively, for a bunch charge of 8.8-aC (16-fC). Interestingly, the measured RMS transverse emittance is determined as 3.8 nm·rad (see Section D and Figure S3 of supplementary material) using knife edge scans of the electron beam at different currents of ML1. The discrepancy between measured (3.8 nm·rad) and simulated (5.6 nm·rad) transverse emittance is most likely due to an overestimation of the mean transverse energy (MTE; ~0.5-eV) in GPT simulations often determined within the severe space-charge regime. In reality, the MTE is lower when operating the instrument in the no-to-low space-charge regime, (i.e., <0.5 eV).

*3. Reciprocal-space resolution and transverse coherence length*

The reciprocal-space resolution and transverse coherence length of our instrument were characterized experimentally by measuring the electron diffraction pattern from a monocrystalline gold film (Plano GmbH, 11-nm) and a polycrystalline aluminium film (Plano GmbH, 31-nm) using a 95-keV electron pulse containing ~100 electrons on-target. These reference samples are often used to characterise both gas-phase and solid-state UED instruments. Fig. 5a shows distinct Bragg diffraction spots from the monocrystalline film. Our instrument was further characterized for isotropic electron diffraction signals, which are typically measured in gas-phase UED, by measuring the diffraction pattern generated from the polycrystalline film, as shown in Fig. 5b. From the FWHM of the first-order diffraction peaks, we obtain a reciprocal-space of 0.063

Å$^{-1}$. An excellent agreement is observed between measured and simulated data, the latter of which is calculated using a powder electron diffraction simulation software (CrystalMaker®)[74]. Furthermore, the RMS transverse coherence length, $\varepsilon_{co}$, of our UED instrument was experimentally characterized using the method described in Ref.[11] that takes into account the reciprocal-space RMS widths of and distance between two closely-lying Bragg electron diffraction signals. Here, we used the electron diffraction signals corresponding to the $(hkl) = (220)$ and $(420)$ Bragg peaks in gold monocrystalline film and the $(hkl) = (220)$ and $(311)$ peaks in polycrystalline aluminium. We obtain a $\varepsilon_{co}$ value of 3.8-nm from both data.

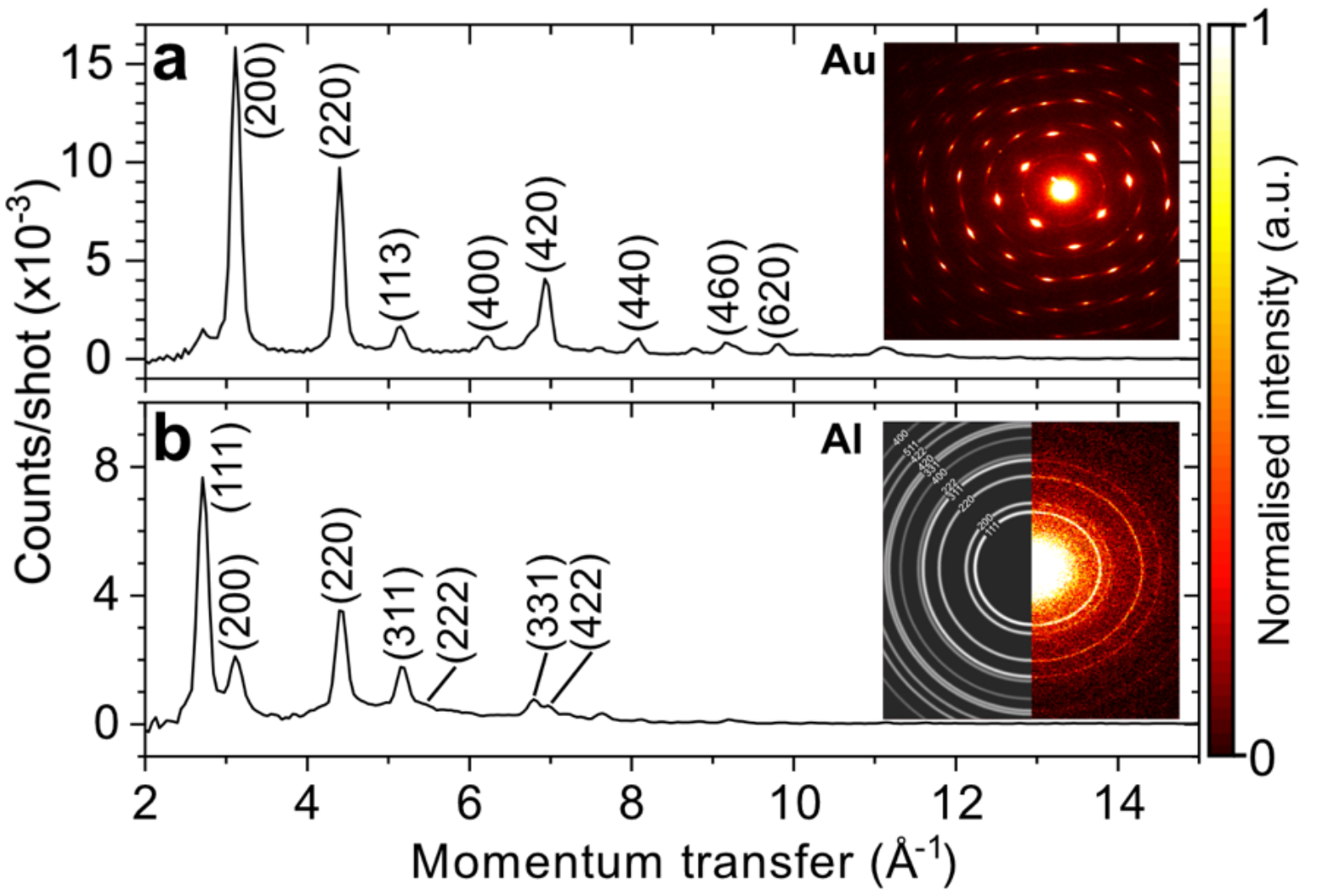

**FIG. 5. (a-b)** Electron diffraction patterns of (a) 11-nm monocrystalline gold (Au) and (b) 31-nm polycrystalline aluminium (Al) measured using a 95-keV electron pulse containing ~100 at 30-kHz with a direct electron detector. Both data were measured without apertures. Insets: measured detector images, with simulated data shown for aluminium.

*4. Electron pulse pointing and intensity jitter*

Fig. 6 shows the long-term pointing and intensity drift of the electron beam measured with a direct electron detector over a two-hour period. The beam pointing and intensity fluctuations of the electron beam were characterized experimentally (see Fig. 6d-f). A Gaussian fit to the histogram distribution of electron intensity in Fig. 6d shows that the electron beam has a 5% (FWHM) intensity fluctuation. While a horizontal and vertical pointing jitter of 23-μm and 28-μm, respectively, are observed (see Fig. 6e-f).

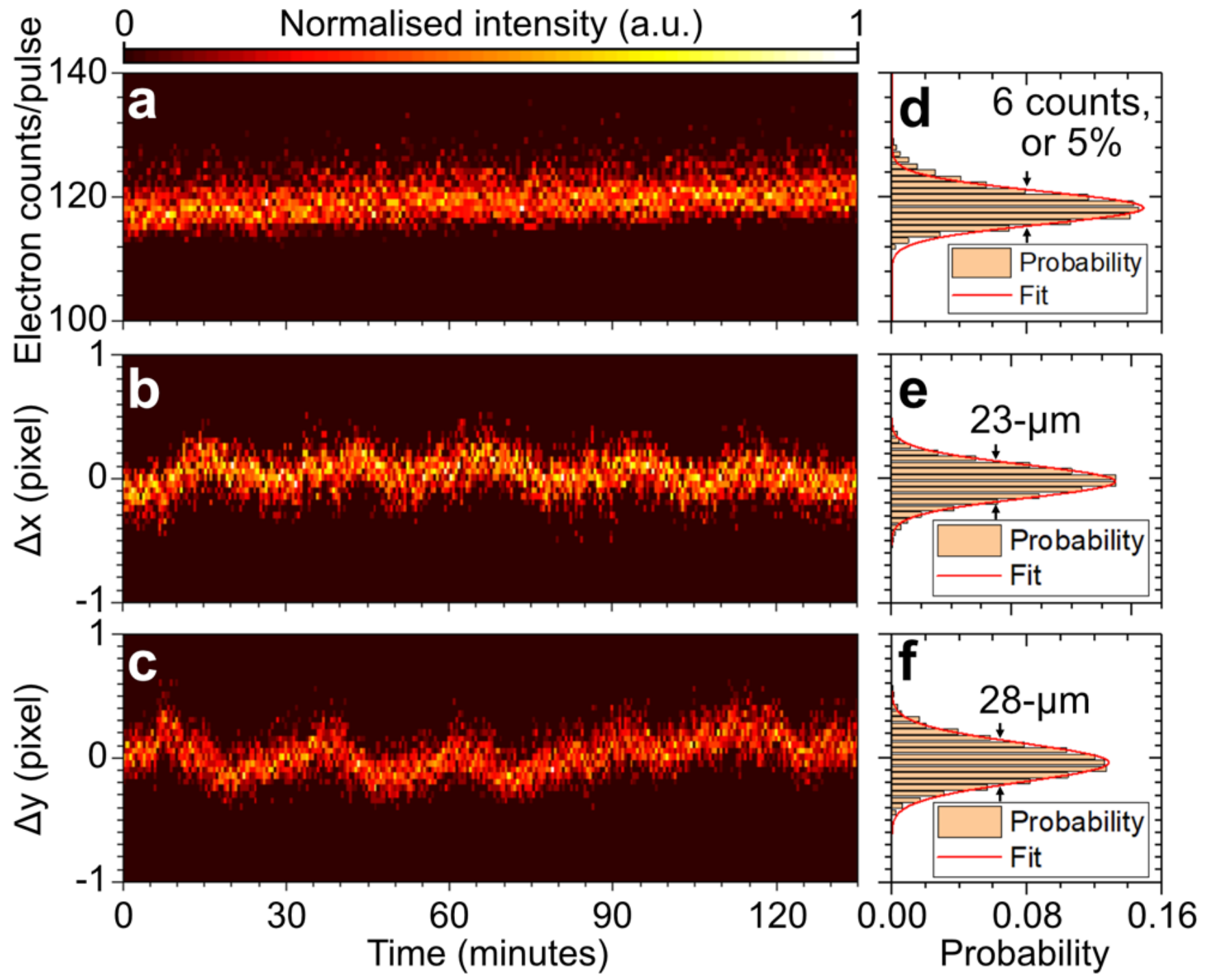

**FIG. 6. (a-c)** Two-dimensional plot of the (a) number of electron counts per pulse, and changes in the (b) $x$-centre position ($\Delta x$), and (c) $y$-centre position ($\Delta y$) as a function of time. A total of ~8,100 images were acquired, with each image capturing the electron beam, which had an average of ~110 electrons, over an exposure time of 1-s. **(d-f)** Histogram plots of the data corresponding to panels (a-c) with a Gaussian fit applied. All values are given in FWHM.

*5. Comparison to other UED instruments*

The transverse diameter and reciprocal-space resolution of the electron beam in our instrument is a factor of two-to-three smaller than other instruments employing bunch charges exceeding 8.8-aC[13,15,19,44]. This reduction is attributed to a lower transverse emittance enabled by the use of an electron beam with a reduced bunch charge (8.8-aC). The measured transverse emittance at 8.8-aC (3.8 nm·rad) is nearly a factor of ten smaller than that reported for most other instruments typically operating at 16-fC (10 nm·rad)[22,23,75]. Moreover, the larger RMS transverse coherence length, $\varepsilon_{co}$, of our UED instrument (3.8-nm) compared to the state-of-the-art (~3-nm)[11,13,76] has important implications for coherently imaging larger lattice unit cell and molecular structures. For example, for solid-state samples, a larger $\varepsilon_{co}$ would mean that the lattice unit cell could be probed an additional number of times, improving the signal-to-noise ratio of the measured electron diffraction signal. Notably, sub-nm·rad emittance, nm-scale transverse electron beam diameters, and coherence lengths of 10 nm or higher have been demonstrated with the next generation of photocathode materials.[22,24,58,61,77] Furthermore, the pointing stability of our instrument (≤28-μm) is similar to that reported for the MeV-UED instrument at SLAC (33-μm)[13].

## B. Direct electron detection

*1. Saturation effects in a direct electron detector*

Figure 7a shows detector images of the unscattered electron beam containing a varied number of electrons, from 10 electrons/pulse to more than 4100 electrons/pulse, recorded after passing through a monocrystalline 11-nm gold thin film sample. Saturation effects become apparent using an electron pulse containing 134 electrons (see Fig. 7a4). For example, the vertical profile of the electron beam (see Fig. 7d) is modified from a Gaussian distribution below saturation to a bimodal distribution when the central pixels become saturated by an electron beam containing 134 electrons or more. Furthermore, in the case of more than 4100 electrons/pulse (see Fig. 7a8), significant saturation effects occur; the central pixels detecting the central portion of the electron beam deviate from a normal counting detector behaviour. We note nevertheless that when the central pixels of the detector are saturated by the unscattered electron beam, the scattered signal observed at larger radii on the detector is not affected, as shown in Fig. 7i. This demonstrates that every pixel operates as an independent electron counter. This is in contrast to the typically-employed CCD[65], EMCCD[13], and CMOS[64] detectors.

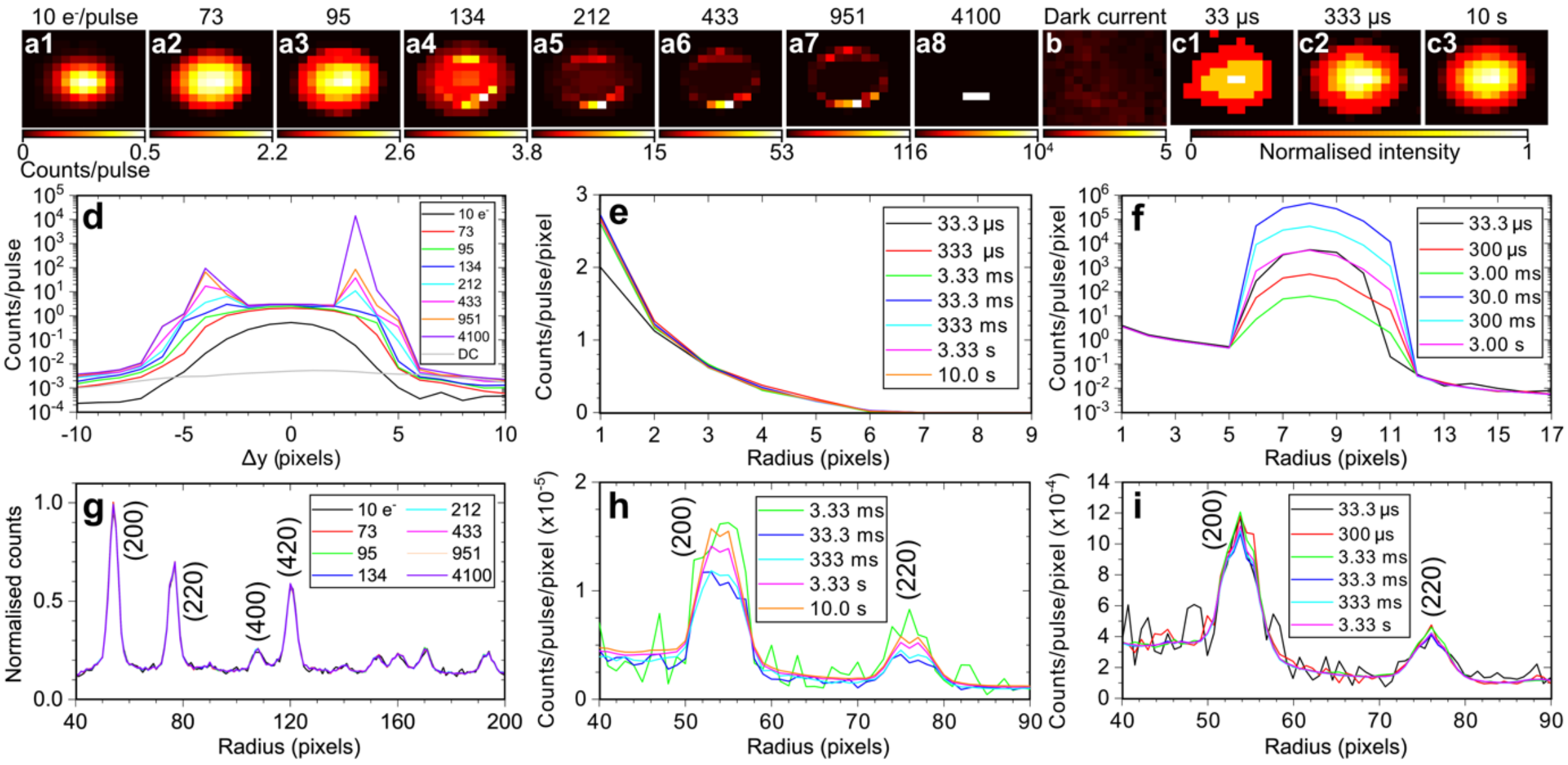

**FIG. 7.** Saturation effects in DECTRIS QUADRO direct electron detector. **(a1-a8)** Detector images measured with a different number of electrons per pulse as indicated at the top of images 1-8. Exposure time of 0.1-s was used. **(b)** Detector image of dark current contribution with a colourmap scaling that was multiplied by a factor of 10 relative to that of image a1. Exposure time of 0.1-s was used. **(c1-c3)** Detector images of an electron beam containing $10^2$ electrons measured with different exposure times. **(d)** Vertical beam profile of the primary unscattered electron beam after interaction with an 11-nm monocrystalline gold sample using an electron pulse containing different numbers of electrons. **(e-f)** Radial distribution of electron beam measured with different exposure times using an electron beam containing (d) $10^2$ electrons and (e) $10^4$ electrons. **(g-i)** Radial distributions of Bragg diffraction peaks from the sample corresponding to data from panels (d-f). Panel (g) was normalised to the sum of the total intensity corresponding to a radius of 30 – 200 pixels for a like-for-like comparison of data measured with different number of electrons. Panels (a,b,d,g) were averaged over 100 images, while all other panels were measured with a single image.

Furthermore, with an exposure time of 33-μs (i.e., 1 shot), the radial distribution of an electron beam containing 100 electrons is measured with a high SNR ratio. While an exposure time of at least 3.33-ms is required to measure the first-order (200) and (220) Bragg diffraction peaks of monocrystalline gold with a SNR ratio of four (see Fig. 7h). Measurement of electron signals significantly above the pixel saturation threshold (see Fig. 7f) reveals that the corresponding pixels behave like a paralyzed counting detector,[78] with signal decreasing as exposure time increases from 33.3-μs to 3.0-ms. Moreover, a sudden intensity surge is observed at an exposure time of 30.0-ms, followed by a decline in signal at longer exposure times. Given the presence of two 16-bit counters in each pixel, this observed saturation behaviour is attributed to the initial paralysis of the first 16-bit counter, followed by the subsequent paralysis of the second 16-bit counter. Notably, in this regime, an exposure time of 33.3-μs (*i.e.*, a single shot) is sufficient to measure the first-order diffraction peaks with an SNR ratio of four (see Fig. 7i), demonstrating the high sensitivity of this detector.

Specifically, each pixel of this detector utilizes a retrigger mechanism which counts the length in time that the amplitude of the measured signal is above a predefined value corresponding to a threshold energy (in our case 12-keV). This length in time is measured as multiples of a programmable time, called the retrigger time (*i.e.*, the time that the signal from one electron is above threshold). The amplitude of the signal is approximately proportional to the energy deposited by the 95-keV electron in that pixel. Therefore, when multiple electrons impinge on the same pixel, this generates a signal with an amplitude that is the sum of the energy deposited by each electron. At every retrigger time that the signal is still above threshold, a count is added to the electron counter. Thus, longer time-over-threshold durations enable the QUADRO to distinguish multiple hits from a single hit.[68,79] Under continuous illumination, a limit of $10^7$ electrons/second/pixel with a retrigger time of 10-ns was established.[68] In our pulsed operation, conducted below saturation, our analysis from Fig. 7d-e indicates that approximately three electrons per pulse can be detected by a single pixel (*i.e.*, $9 \times 10^4$ electrons/second/pixel, and a retrigger time of 825-ns). Thus, the QUADRO's retrigger feature enables the counting of up to three electrons arriving within the electron pulse duration (<350-fs) in a single pixel at 30-kHz.

*2. Shot-to-shot jitters and their correction*

We have characterized the electron beam's intensity and pointing fluctuations on a shot-to-shot basis. To demonstrate the capability to correct for such fluctuations, we performed measurements immediately after applying current to the two magnetic solenoid lenses, without allowing time for thermalisation, to ensure maximum fluctuation in the electron beam. Employing a similar histogram analysis as that presented earlier, we find that an electron beam containing 113 electrons exhibited a FWHM intensity jitter of 25.7 electrons or 22.7%, as shown in Figure 8a. This is approximately equivalent to the shot noise limit assuming Poisson statistics (~22.2%)[64]. Furthermore, changes in the electron beam's centre position, $\Delta x$ and $\Delta y$, were 56-μm and 75-μm, respectively, as shown in Figure 8b-c.

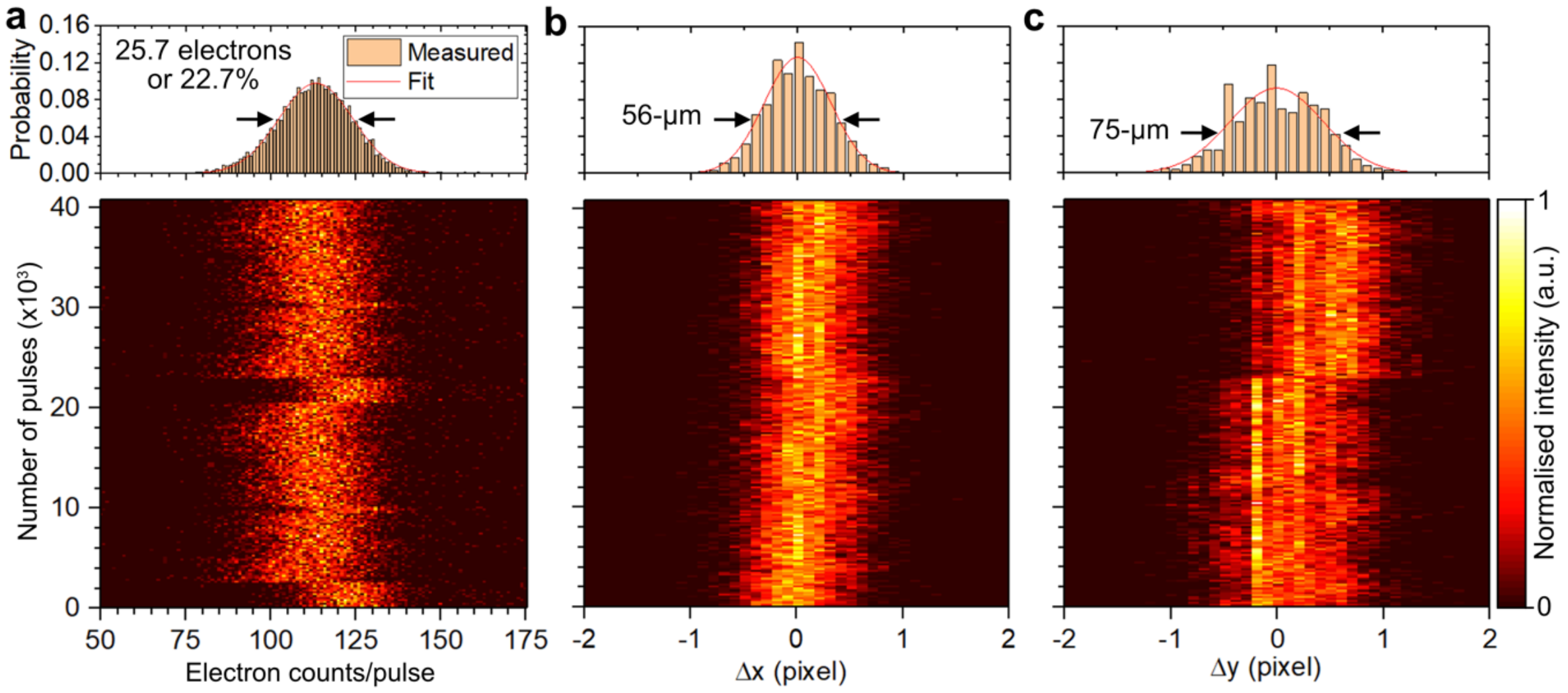

**FIG. 8.** Shot-to-shot analysis. **(a)** Two-dimensional plot of the number of electrons per shot as a function of number of pulses. An electron beam containing an average of 113 electrons was measured with an exposure time of 33.3-μs. A total of 40,000 images were recorded. A histogram plot of the associated data is shown at the top of the panel with a Gaussian fit applied. **(b-c)** Same as in panel (a) but for changes in the $x$ (b) and $y$ (c) centre positions of the electron beam.

An algorithm was developed that implements multiple corrections on a shot-to-shot basis, applicable to data measured with any direct electron detector. Initially, each element within the 512 × 512 pixel matrix of every image was normalised by the total number of electrons present in the unscattered beam. The resulting sum detector image before and after intensity correction is shown in Fig. 9a and 9b, respectively.

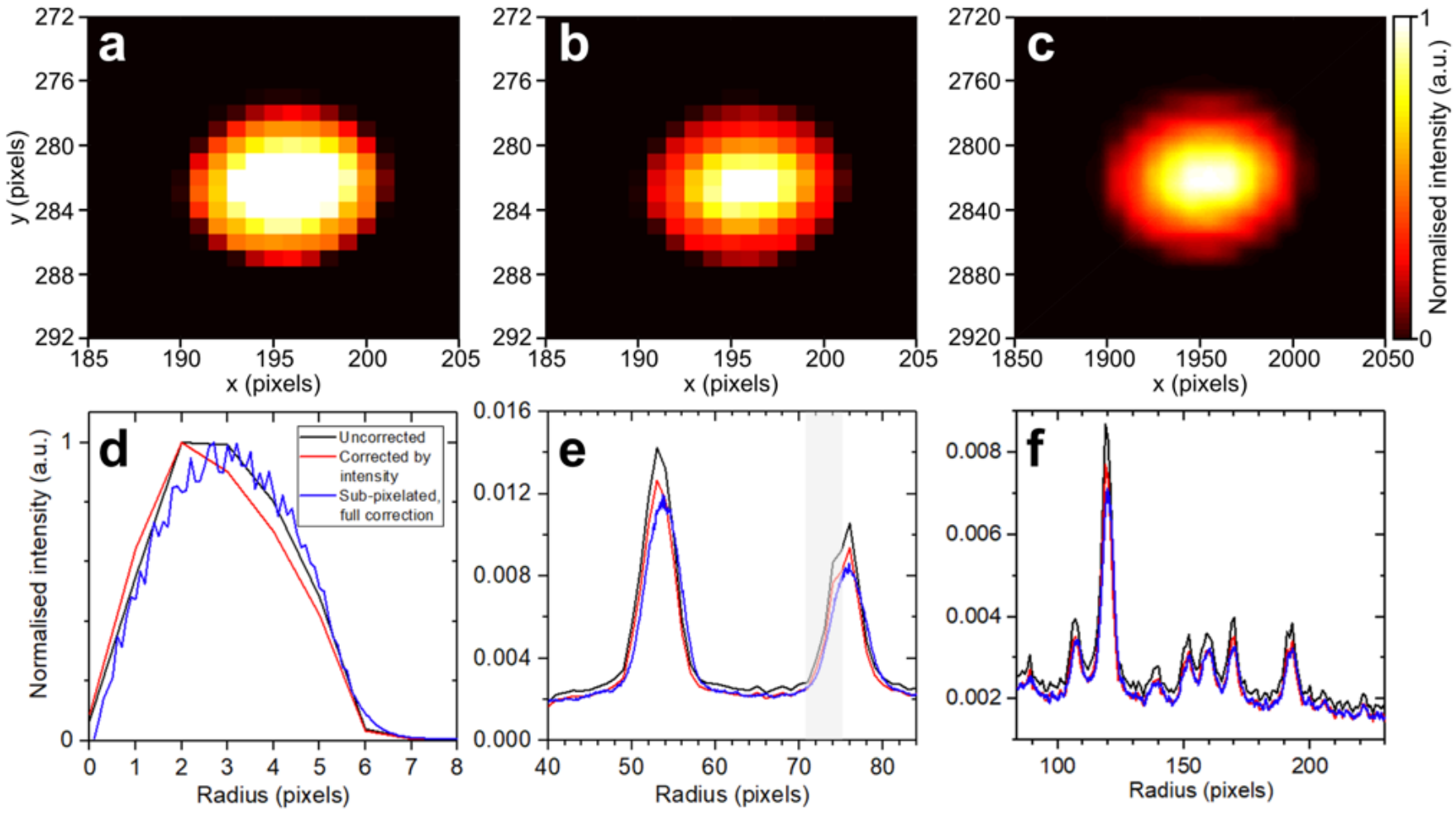

**FIG. 9.** Shot-to-shot correction of electron beam pointing and intensity jitter with sub-pixel accuracy. **(a-c)** Summed detector image of electron beam (a) before any correction, (b) after intensity jitter correction, and (c) after intensity and pointing jitter correction with sub-pixelation of factor 10. The images shown were summed over ~40,000 shots corresponding to the data shown in Figure 8. **(d)** Radial distribution of primary unscattered electron beam before and after correction. **(e-f)** Radial distribution of Bragg diffraction peaks from 11-nm monocrystalline gold before and after correction.

Exploiting the ability to bin $\Delta x$ and $\Delta y$ into 0.1-pixel bins (see histograms at top of Fig. 8b-c), the dimensions of the detector image were expanded from the original 512 × 512 pixels to 5120 × 5120 pixels by sub-dividing each pixel into a 10 × 10 grid.

This is referred to as the sub-pixelation method. Each element within this 10 × 10 pixel array retained the same intensity value as its corresponding original pixel. Changes in centre positions were corrected on a shot-to-shot basis with 0.1-pixel precision of the original 512 × 512 pixel matrix, utilizing the newly formed 5120 × 5120 pixel matrix. The resulting sum detector image of the fully corrected electron beam is depicted in Fig. 9c. The normalised radial distributions before and after correction are presented in Fig. 9d-f for both unscattered and scattered electrons. It is evident that a more uniform unscattered electron beam is achieved after full correction (blue line in Fig. 9d) compared to the intensity-only corrected data (red line) and the uncorrected data (black line). Notably, oscillations in the fully corrected signal (see Fig. 9d) are due to aliasing effects arising from the use of 0.1-pixel bin sizes during the sub-pixelation process. Moreover, the shoulder observed for the (220) Brag peak in the uncorrected data is no longer present in the fully corrected dataset (see grey shaded area in Fig. 9e). Such a shot-to-shot correction offers the capability to correct measured scattering data despite intensity and pointing fluctuations in the primary electron beam. This capability has not been attainable thus far in UED instruments utilizing other types of electron detectors mentioned previously. In the subsequent section, the efficacy of correcting the electron beam intensity jitter is demonstrated.

### C. Time-overlap between optical-pump and electron-probe

We investigate the time-resolved pump-probe capability of our UED instrument through space-charge deflection of the primary electron beam (213-fs FWHM simulated, 35 electrons/pulse, 102-μm FWHM diameter at sample) by photoelectrons generated at the surface of a meshgrid (300 lines/inch) using an 800-nm optical pump pulse (60-fs FWHM, 180-μm FWHM, 13.6 mJ/cm$^2$).[11] Figure 10a displays the detector image of the electron beam measured at a pump-probe delay of -15 ps. Figure 10b presents the difference images of electron beam measured at positive delays, obtained by subtracting the image at -15 ps each image measured at positive delays. These difference images clearly show that the electron beam is vertically deflected (see blue arrow) due by the photoelectrons generated at the surface of the meshgrid.

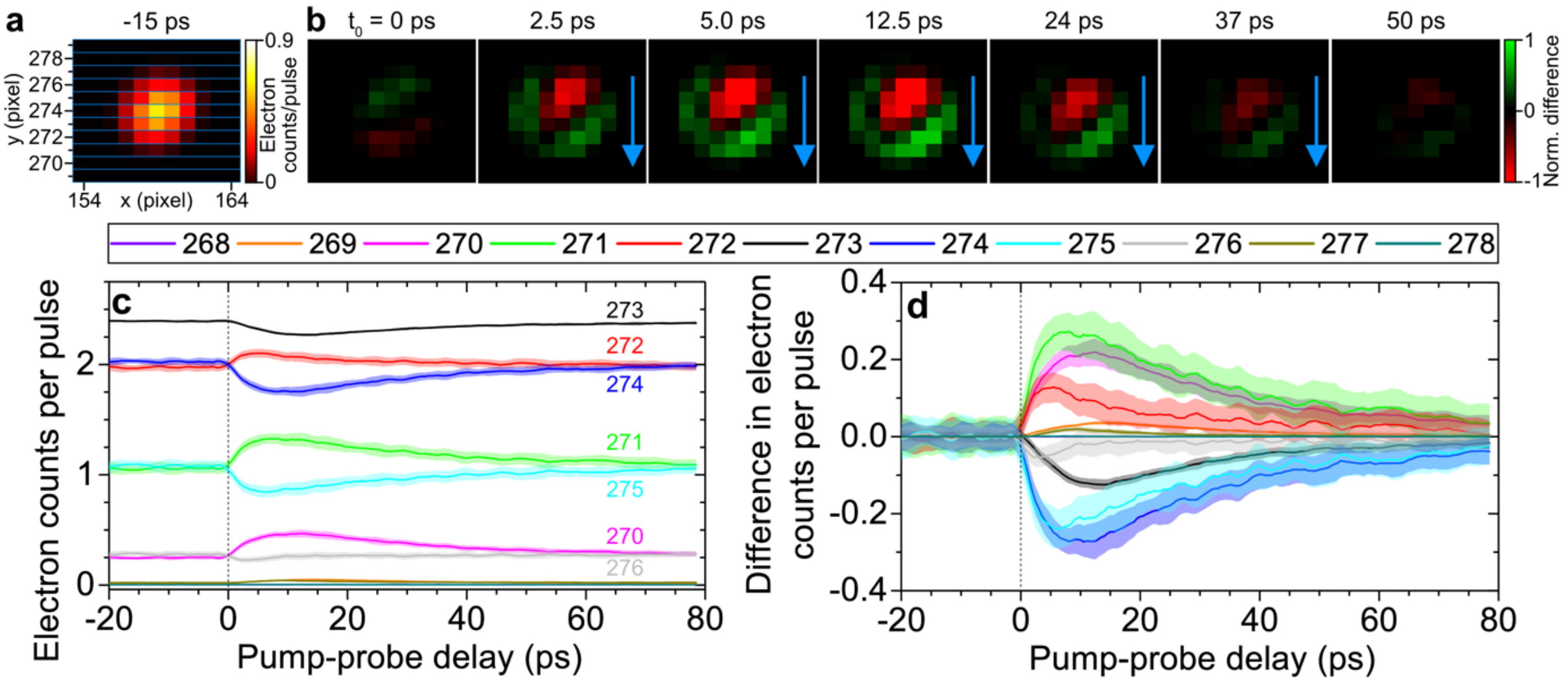

**FIG. 10.** Finding time-zero overlap between the optical-pump and electron-probe pulses by space-charge deflection of the electron beam containing 35 electrons. **(a)** Detector image of an electron pulse, measured at a pump-probe delay of -15 ps, passing through a copper meshgrid (300 lines/inch) with a 100-μm copper aperture placed in-front of the meshgrid. An optical pump pulse was focussed to a ~100-μm diameter spot on the meshgrid with a fluence of 13.6 mJ/cm$^2$. **(b)** Normalised difference images at positive pump-probe delays, obtained by subtracting the background image at -15 ps from panel (a). Direction of space-charge deflection of electron beam is shown by blue arrows. **(c)** Electron counts per shot integrated across different horizontal strips of the detector image corresponding to a specific $y$-pixel integrated over a fixed $x$-pixel range of 154-164 (see blue shaded rectangles in panel (a) at -15 ps). **(d)** Same as in panel (c) but with the average counts at negative delays subtracted in each $y$-pixel horizontal strip.

To quantify the measured time-resolved signal, we integrate the electron counts in each row of $y$-pixels across all $x$-pixels, effectively creating a horizontal strip detector (depicted by blue rectangles in Fig. 10a at the delay of -15 ps). For each pump-probe delay ($\Delta t$), we subtract the pump-probe signal ($I_{\mathrm{pp}}$; see Fig. 10b) by a reference probe-only signal ($I_0$) which provides the difference electron signal ($\Delta I$; see Fig. 10c). The $I_0$ signal is a single image generated from the average of all images measured at negative $\Delta t$ values. Figures 10c-d reveal a substantial increase in electron counts along the $y = 270 - 272$ pixel row strip lines (see green lines). This is accompanied by a decrease in signal along the $y = 273 - 276$ pixel rows (see light-blue lines) in the first 10-ps after $t_0$. These variations in electron signal correspond to the vertical deflection of the electron beam by photoelectrons, slowly returning to its original position by approximately 80 ps.

Figure 11a illustrates the fluence dependence of the time-resolved difference contrast signal, $\Delta I/I_0$, arising from space-charge deflection. We define a circular region of interest (ROI) with a radius of four pixels that is centred on the electron beam. On a pixel-by-pixel basis, and at each pump-probe step, we normalise the difference electron signal, $\Delta I$, by the sum of the electron beam to correct for fluctuations in the electron beam intensity. The $\Delta I$ signal in each pixel of the ROI is further normalised by $I_0$ to derive the difference contrast signal, $\Delta I/I_0$. A minimum fluence of 11 mJ/cm$^2$ is necessary to detect a discernible $\Delta I/I_0$ signal. As the fluence increases, so does the $\Delta I/I_0$ signal, reaching up to $3 \times 10^{-3}$ (*i.e.*, a 0.3% change relative to the probe-only signal) with a fluence of 76 mJ/cm$^2$.

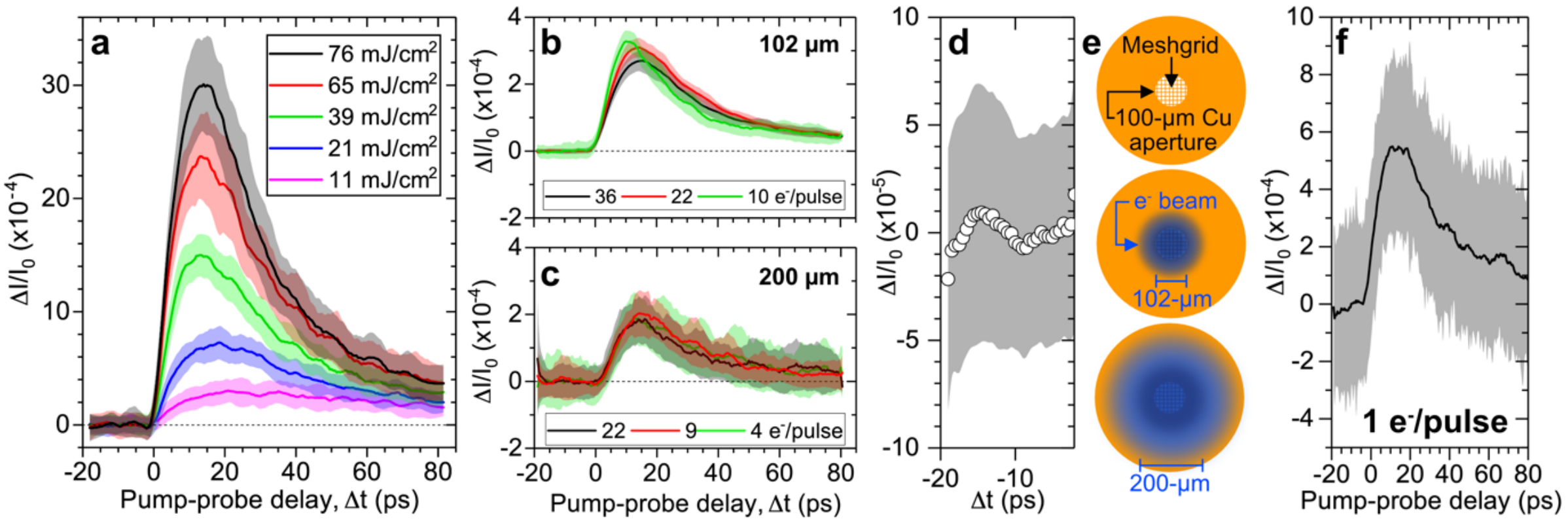

**FIG. 11. (a)** Difference in pump-probe signal relative to the reference probe-only signal, $\Delta I/I_0$, as a function of pump-probe delay at different pump fluences. The $\Delta I/I_0$ signal was integrated across a circular region of interest with radius of four pixels, and was corrected for fluctuations in electron beam intensity. At all fluences, an electron beam containing 140 electrons had 40 electrons passing through a 100-μm aperture placed in front of the meshgrid sample. **(b-c)** $\Delta I/I_0$ as a function of pump-probe delay for an electron beam with varying numbers of electrons but with two different FWHM transverse diameters of (b) 102-μm (solid distributions)

and (c) 200-μm (dashed distributions) at the sample position. The legend indicates the number of electrons in the beam that passed through the 100-μm aperture. **(d)** $\Delta I/I_0$ at negative pump-probe delays corresponding to the red distribution in panel (b). **(e)** Schematic of the aperture, meshgrid and electron beam size. **(f)** $\Delta I/I_0$ as a function of pump-probe delay measured with an electron pulse containing 1 electron before the aperture. The $\Delta I/I_0$ signals of panels (b-d,f) were measured at a constant pump fluence of 13.6 mJ/cm$^2$.

In our investigation, we also study the influence of the electron beam's transverse diameter at the sample position on the measured $\Delta I/I_0$ signal. Figure 11b-c shows measurements conducted with two different electron beam FWHM diameters: 102-μm (Fig. 11b) and 200-μm (Fig. 11c). In both cases, increasing the number of electrons passing through the 100-μm aperture placed in-front of the meshgrid leads to no significant change in the $\Delta I/I_0$ signal when considering their corresponding error bars. However, employing a smaller electron beam diameter at the sample (see Fig. 11e for a schematic) yields a $\Delta I/I_0$ signal with approximately 50% higher contrast. Our results complement previous studies utilizing similar space-charge deflection techniques. Previous investigations often employed electron beams in either shadow imaging mode[80,81], characterized by a notably larger beam diameter at the sample, or in an intermediary mode between shadow and reciprocal-space imaging, using a moderately large beam diameter. Our study demonstrates that achieving a measurable $\Delta I/I_0$ signal is equally feasible in reciprocal-space imaging mode. This is accomplished by employing an electron beam with the smallest FWHM diameter, which in our case was equivalent to the aperture's inner diameter. Fig. 11d shows the $\Delta I/I_0$ signal at negative pump-probe delays of the red distribution shown in Fig. 11b, where the average of the absolute value (standard deviation) of $\Delta I/I_0$ is $5 \times 10^{-6}$ $(5 \times 10^{-5})$. This represents a 400-fold (40-fold) reduction compared to the minimum $\Delta I/I_0$ measurable with EMCCD detection using the MeV-UED instrument at SLAC $(2 \times 10^{-3}$ or 0.2%). The SNR ratio observed in Fig. 11b-c ranges from 60 to 40, which starkly contrasts the SNR ratio of 5 to 2.5 typically obtained with EMCCD detection[13]. The significant disparity in SNR ratios emphasizes the critical importance of acquiring electron signals with minimal noise, which is inherently possible with direct electron detection.[82] For example, the in-pixel digital counting capability of a direct electron detector minimizes inherent sources of noise in the pixel (*e.g.*, gain, integration operations) and readout electronics which have often limited the SNR ratio in charge-integrating analog detectors,[69] such as CCD and CMOS sensors. Our HiRep-UED instrument employing direct electron detection is primarily limited by shot noise[64], arising from the measured signal statistics, and source noise[64], resulting from fluctuations in the electron beam.

To further demonstrate the capabilities of our HiRep-UED instrument, we repeated the time-resolved measurements using an electron pulse containing a single electron, measured before the aperture. Figure 11f shows similar trends in the time-resolved signal as discussed earlier. Notably, time-resolved signals obtained using 1 electron/pulse (before aperture) are measurable at a detected count rate of 0.03 electrons/pulse (after aperture). This highlights the capability of direct electron detection to measure time-resolved signals with low SNR at very low count rates.

Additionally, the pulse duration of 1 electron/pulse in our system is simulated to be 174-fs (FWHM) at the sample position. Since the electron pulse is temporally uncompressed, the jitter between the optical pump pulse and electron probe pulse is negligible. Moreover, the velocity mismatch with our 95-keV beam is also negligible for the (<100-nm thick) solid films of samples studied here. Therefore, the main contributions to the IRF are from the electron probe pulse duration, $t_e$ (174-fs), and the optical pump pulse duration, $t_p$ (60-fs). Consequently, the lowest IRF of our HiRep-UED instrument is determined to be 184-fs (FWHM) for 1 electron/pulse, as given by

$$IRF = \sqrt{t_e^2 + t_p^2}.$$

### D. Time-resolved dynamics in photoexcited aluminium thin film

We next demonstrate the ability to measure time-resolved elastic electron scattering signals from a 31-nm thin film of polycrystalline aluminium as a prototypical system for expected isotropic gas-phase electron scattering signals. The aluminium sample is optically excited with an 800-nm pulse (<60-fs FWHM, ~180-μm FWHM diameter, 2 mJ/cm$^2$) and probed by a 95-keV electron pulse containing 134 electrons (313-fs FWHM simulated, ~100-μm FWHM diameter). The expected IRF is 319-fs (FWHM). Figure 12a shows the $\Delta I/I_0$ signal as a function of momentum transfer and pump-probe delay, while Figure 12b presents the $\Delta I/I_0$ signal integrated over pump-probe delays greater than +1 ps.

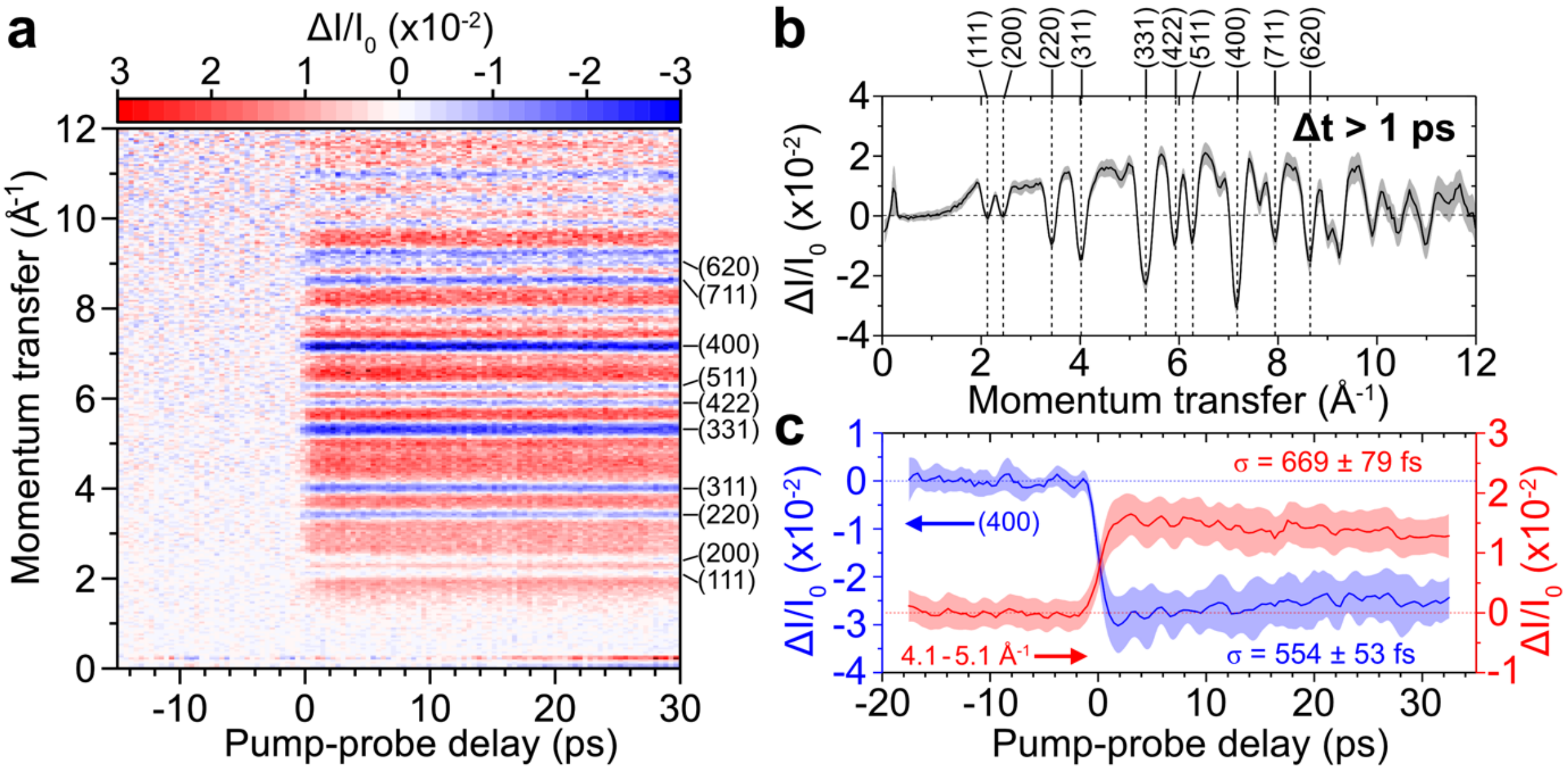

**FIG. 12. (a)** Measured $\Delta I/I_0$ signal of 31-nm polycrystalline aluminium thin film optically excited by an 800-nm pump pulse with a fluence of 2 mJ/cm$^2$ as a function of momentum transfer and pump-probe delay. Diffraction peaks are labelled. **(b)** $\Delta I/I_0$ as a function of momentum transfer integrated at pump-probe delays of greater than +1 ps. **(c)** $\Delta I/I_0$ as a function of pump-probe delay of the (400) diffraction peak and the $\Delta I/I_0$ signal between 4.1 – 5.1 Å$^{-1}$. An electron pulse containing 134 electrons before a 200-μm aperture was used.

The first two diffraction peaks of (111) and (200) show no appreciable change in $\Delta I/I_0$. While diffraction peaks at higher momentum transfers, from (220) to (620) and higher, show a decrease in $\Delta I/I_0$. For example, the $\Delta I/I_0$ signal of the (400) diffraction peak decreases by approximately 3% with a time constant of $\sigma = 554 \pm 53$ fs, as shown in Figure 12c. This step-function decrease in the $\Delta I/I_0$ signal of aluminium is attributed to a transient thermal Debye-Waller effect[83,84].

Interestingly, electron scattering signals measured at momentum transfers between diffraction peaks, referred to as the diffuse background, show a positive change in $\Delta I/I_0$. For example, the integrated $\Delta I/I_0$ signal between 4.1 – 5.1 Å$^{-1}$ shows an increase of 1.5% in signal with a time constant of $\sigma = 669 \pm 79$ fs, as shown in Figure 12c. The diffuse background signal increases simultaneously with the decrease in higher-order diffraction signals. A comparable positive $\Delta I/I_0$ signal at positive delays was reported by Siwick *et al.*[33] during the non-reversible solid-to-liquid phase transition of thin film aluminium at 70 mJ/cm$^2$, which required the translation of the sample. In contrast, the data shown in Figure 12 is composed of multiple pump-probe scans at a much lower fluence of 2 mJ/cm$^2$, all measured without translating the sample. At a similar fluence of approximately 2 mJ/cm$^2$, previous studies have reported the presence of coherent acoustic phonon modes in thin film aluminium[83–85] following photoexcitation by an 800-nm pump pulse. These modes were detected as shifts in the centre position of the diffraction peak. In our measurements, however, we do not observe such oscillations in the diffraction peaks.

A comparable increase in the thermal diffuse scattering signal was observed in thin bismuth films by Sokolowski-Tinten *et al.*[86], occurring simultaneously with a decrease in the intensity of diffraction peaks. The timescales of these changes were also similar. The observed time-resolved changes in bismuth were attributed to a transient Debye-Waller effect, which caused an increase in the atomic displacement after sample excitation, subsequently leading to the diffuse scattering from phonons. Given the similar trend observed in the aluminium data shown in Figure 12, we attribute the measured time-resolved changes to a similar transient Debye-Waller effect that results in diffuse scattering from phonons. The timescale of these time-resolved changes occurs more than twice faster in aluminium ($\sigma \sim 600$ fs) compared to bismuth ($\sigma \sim 1500$ fs). This faster timescale can be attributed to stronger electron-phonon coupling in aluminium, which facilitates a more rapid cooling of the hot electrons and a faster equilibration with the lattice in the film.

### E. Simulation of temporal compression in the HiRep-UED instrument

In this work so far, we have utilized an electron pulse without temporal compression. However, ongoing efforts are directed towards achieving temporal compression of the electron pulse using an RF compression scheme. While the RF compression of our electron pulse is subject to a future publication, it is still pertinent to explore the anticipated capabilities of our instrument when employing RF-compressed electron beams. Fig. 13 shows the FWHM pulse duration and transverse beam diameter at the sample position predicted by GPT simulations for an RF-compressed (red squares) and uncompressed (black circles) electron beam containing from one to $10^7$ electrons.

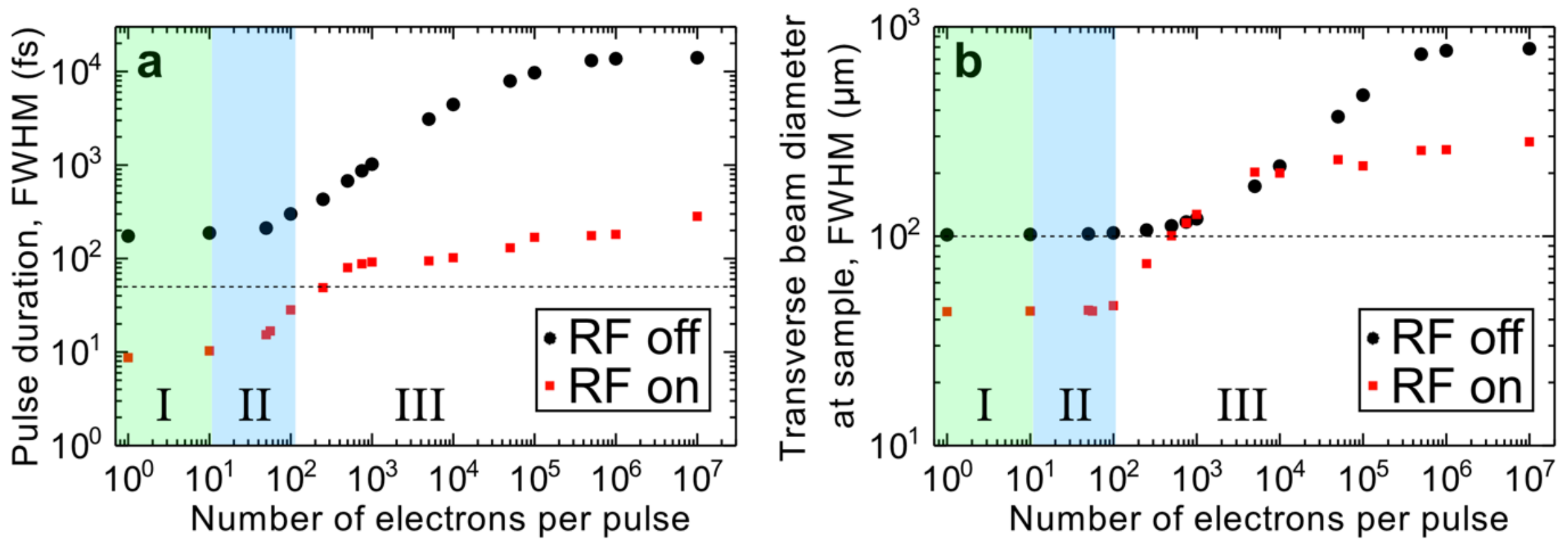

**FIG. 13. (a-b)** General particle tracer (GPT) simulations of the (a) pulse duration and (b) transverse beam diameter of an electron pulse containing a varied number of electrons at the sample. All values are given in FWHM. Simulated data are shown with the radiofrequency (RF) cavity on (black circles) and off (red squares) where the electron beam is temporally compressed and uncompressed, respectively. The UV pulse duration used in the simulations was 90-fs (60-fs) FWHM with the RF cavity on (off). Three regions of space-charge are indicated: (I) no-to-low space-charge (green shaded), (II) low-to-moderate space-charge (blue shaded), and (III) severe space-charge dispersion. Horizontal dashed lines indicate thresholds for an electron beam with a 50-fs duration and a 100-μm transverse diameter at the sample.

Here, a UV pulse duration of 60-fs (90-fs) FWHM was used in the simulations with (without) RF-compression. The simulations show three distinct regimes of space-charge effects across the single-electron to single-shot operating regimes: no-to-low space-charge (regime I; green shaded), low-to-moderate space-charge (regime II; blue shaded), and severe space-charge (regime III). By implementing RF compression in our instrument, we can achieve pulse durations of 50-fs or less (see dashed line in Fig. 13a) with electron beams containing 250 electrons or fewer, as predicted by GPT simulations. This represents a more than tenfold reduction in pulse duration compared to uncompressed electron beams with the same bunch charge. Furthermore, utilizing electron beams containing 250 electrons at 30-kHz results in an electron flux of $7.5 \times 10^6$ electrons/second, which is more than two times higher than that of the MeV-UED instrument at SLAC operating at 0.36-kHz[13]. The predicted 50-fs compressed pulse duration would be over two (nearly four) times shorter than that of the MeV-UED instrument at SLAC[13] (keV-UED set-up in Ref.[15]). Such a short pulse would enable imaging of nuclear dynamics in gas-phase photochemical reactions that reach completion in less than 500-fs,[87–89] which have thus far been too rapid to image with existing UED instruments. Moreover, ensuring a minimal transverse diameter of the electron is also crucial to minimize the pump pulse diameter and average power requirements of the laser system operating at high repetition rates. With RF compression, it is anticipated that an electron beam diameter of 100-μm (see dashed line in Fig. 13b) could be achieved with a beam containing 500 electrons. In general, GPT simulations predict that electron beams containing 1 - $10^7$ electrons will exhibit compressed (uncompressed) pulse durations of 9 fs – 283 fs (174 fs – 14 ps) and transverse diameters of 44 μm – 282 μm (102 μm – 787 μm). Figure 1 illustrates the range of operating parameters anticipated for our instrument based on GPT simulations (see red dashed lines). These operating regimes are broadly categorized as the ultrashort and high brightness modes. The ultrashort mode is projected to achieve pulse durations of 50-fs or less, while the high brightness mode is capable of pulse durations of below 300 fs but with a substantially higher average current of nearly 70-nA. This average beam current surpasses that of the brightness existing ultrashort keV[44,90] (MeV[13,17]) electron beam sources by more than one (four) order(s) of magnitude.

## V. CONCLUSIONS

In summary, we introduce a novel high repetition rate UED (HiRep-UED) instrument operating at 30-kHz with direct electron detection. We demonstrate the feasibility of operating in the low-to-moderate space-charge regime at 30-kHz, enabling the acquisition of statistically significant electron signals within a reasonable acquisition time of between a few minutes to up to 90 minutes. Within this regime, the electron beam exhibits relatively low transverse and longitudinal emittance, facilitating the use of temporally uncompressed electron pulses containing 1-140 electrons. This setup allows the measurement of time-resolved signals from photoexcited samples with an instrument response function as low as ~184-fs (FWHM). Additionally, transverse focussing of the electron beam to a spot diameter as small as ~100-μm (FHWM) is possible at the sample position. By employing direct electron detection, we demonstrate the ability to measure time-resolved effects in thin film aluminium with a difference contrast $\Delta I/I_0$ signal on the $10^{-5}$ order of magnitude. This high detection sensitivity is made possible by the direct detection of electrons with reduced noise associated with pixel (*e.g.*, gain, integration operations) and readout electronics. Furthermore, direct electron detection enables the measurement of the unscattered primary electron beam, a capability that proves invaluable in correcting for fluctuations in the electron beam's intensity through our experiments under varied experimental conditions.

Our ongoing efforts to improve our setup include the implementation of temporal compression of our electron beam using RF fields generated by a microwave cavity. Firstly, the UV pulse will be chirp-compensated to sub-60-fs using chirped mirrors, suitable for the UV range, possessing negative GDD. Furthermore, the installation of this cavity is complete, and work is currently underway to optimize its functionality. Additionally, we have implemented an active RF-laser synchronization system based on Ref.[43] to correct fluctuations in the RF-laser timing jitter. The implementation of RF compression into our setup is predicted to extend our capabilities, offering a compressed electron beam with variable duration (9 fs to 283 fs) and average current (5-fA to 50-nA). This broad operational range will enable our setup to operate in or between ultrashort and ultrabright modes, catering to diverse experimental requirements. Depending on the timescale of the photo-induced dynamics in gas-phase molecules, the instrument is anticipated to be operated in the ultrashort mode, with an anticipated IRF of less than 100 fs, suitable for studying photochemical reactions completed within 500 fs,[87–89] or in the ultrabright mode, with up to $10^7$ electrons/pulse, ideal for investigating dynamics on the picosecond or longer timescale, or somewhere between these two modes.

Future iterations of UED setups could benefit from several enhancements. One avenue for improvement involves minimizing the excess transverse momentum spread and thereby reducing the transverse emittance of the electron beam. This would then further reduce the transverse electron beam diameter at the sample position. This could be achieved by adopting new photocathode materials with lower work functions[22,91] than traditional options like copper[67] and gold[34]. Moreover, this would enable the use of visible optical pulse in electron photoemission[35,91], allowing to match the photocathode work function[35], resulting in the generation of photoelectrons with minimal excess energy and lower emittance. This approach, facilitated by ultrashort visible pulses, represents a highly desired departure from the commonly used ultraviolet pulses. Another important aspect is increasing the acceleration field strength at the photocathode surface to above the typically achieved 10 MV/m in keV-scale UED instruments.[92,93] While fields exceeding 10 MV/m can produce shorter electron pulses, they also lead to more significant emittance growth at these higher field strengths.[94,95]

Furthermore, our system's current repetition rate of 30-kHz is already well-placed for gas-phase UED measurements planned in the near future. However, scaling the repetition rate of the system to, for example, 100-kHz is feasible using currently available high-average power, femtosecond laser systems. Given that the average power and repetition rate of femtosecond laser systems have continually increased over the last three decades, scaling to hundreds of kHz (or even to the MHz level) is conceivable in the future. Increasing the repetition rate would offer significant benefits, including the ability to utilize electron beams with lower bunch charges. Gas-phase UED measurements could then utilize electron beams experiencing no-to-low space-charge dispersion[10,12,46,50] and optimal emittance properties but at the hundreds of kHz or 1-MHz repetition rate, particularly when combined with direct electron detection. Additionally, ultrashort electron beams with optimal emittance properties offer promising applications in other areas, such as electron energy loss spectroscopy[23,96] with the use of streaking fields[97–99], dipole magnets[100,101] and monochromation techniques[102].

## SUPPLEMENTARY MATERIAL

Further details of the UV-photocathode alignment procedure, HiRep-UED instrument, general particle tracer simulation of UV pulse duration, and transverse emittance, coherence and beam diameter of electron beam are given in the supplementary material.

## ACKNOWLEDGEMENTS

The authors are grateful for fruitful discussions and valuable support provided by Arnaud Rouzée, Brad Siwick, Jom Luiten, Christoph Reiter, Roman Peslin, Wolfgang Krueger, Rainer Schumann, Octave Grob, Tancrede Esnouf, Laurenz Kremeyer, Anshul Kogar, Alexander Hume Reid, and Ming-Fu Lin. We thank DECTRIS (Luca Piazza, Pietro Zambon, Matthias Meffert) for the provision of the QUADRO direct electron detector and valuable discussions. We thank DrXWorks (Jim Franssen, Thomas Luiten), Pulsar Technologies (Bas van der Geer, Marieke de Loos), Light Conversion (Ignas Abromavičius, Valdas Maslinkas) and TOPAG (Hannes Stroebel, Igor Quint) for extensive support. The authors are also grateful for financial support from Max-Born-Institut.

## DATA AVAILABILITY

The data that support the findings of this study are available from the corresponding author upon reasonable request.

## Supplementary Material

### A. UV-photocathode alignment procedure

As a first alignment step, the UV lens was removed and the aperture of an iris positioned before the first piezo-mounted mirror was reduced to a <1-mm diameter size. Using the maximum UV pulse energy available (20-nJ), the unfocussed UV beam impinged on the 1-mm flat part of the photocathode only when the outcoupled UV beam reflected by the second, rectangular in-vacuum UV mirror shows the machining grooves of the photocathode (see Fig. S1a). Moreover, an electron pulse could be measured with the electron detector without the UV lens inserted. To correctly align the UV pulse onto the flat part of the photocathode, the input and outcoupled UV beam crossed each other using the two fixed in-vacuum UV mirrors of the electron accelerator set-up (formerly AccTec BV, now known as Doctor X Works BV).

As a second alignment step, the UV lens was re-inserted and aligned along the optical axis such that there was no coma or astigmatism present in the UV beam profile at the focus. The last piezo-mounted mirror and the UV lens were both mounted to their own respective linear stage, enabling fine adjustment to achieve optimal spatial alignment perpendicular to the optical axis. At the optimal alignment, the brightest outcoupled UV beam was reflected by the second, rectangular in-vacuum UV mirror, which also exhibited machining grooves of the photocathode (see Fig. S1b). A systematic examination of the unfocussed electron beam profile was conducted with the detector positioned as close as feasible to the photocathode (photocathode-detector distance of ~0.5-m). An optimal electron beam profile was determined by translating the last piezo-mounted UV mirror across a $2 \times 2$ mm$^2$ area of the photocathode central flat region. The second last piezo-mounted UV mirror was then adjusted by a relatively large step, and the two-dimensional scan across a $2 \times 2$ mm$^2$ area of the photocathode repeated with the last piezo-mounted mirror. This overall alignment procedure was repeated until the smallest electron beam diameter was obtained which remained consistently small across different positions of the 30-μm UV beam on the 1-mm flat portion of the photocathode. Furthermore, optimization of the UV lens position along the optical axis and adjustment of the distance between the two curved UV mirrors were performed to obtain a symmetrical and minimized electron beam profile.

Precise spatial alignment of the 800-nm optical pump and electron probe pulses was ensured by utilizing a 100-μm diameter copper aperture placed in-front of a meshgrid sample. Optimal alignment is achieved by optimizing the number of detected electrons (nine electrons passing through the aperture) and obtaining a symmetrical image of the meshgrid on a near-infrared sensitive card when the optical pulse passes through the aperture (see image in Fig. S1c measured with a camera). This alignment process utilizes an in-vacuum piezo-driven mirror to fine-tune the optical pulse alignment through the aperture.

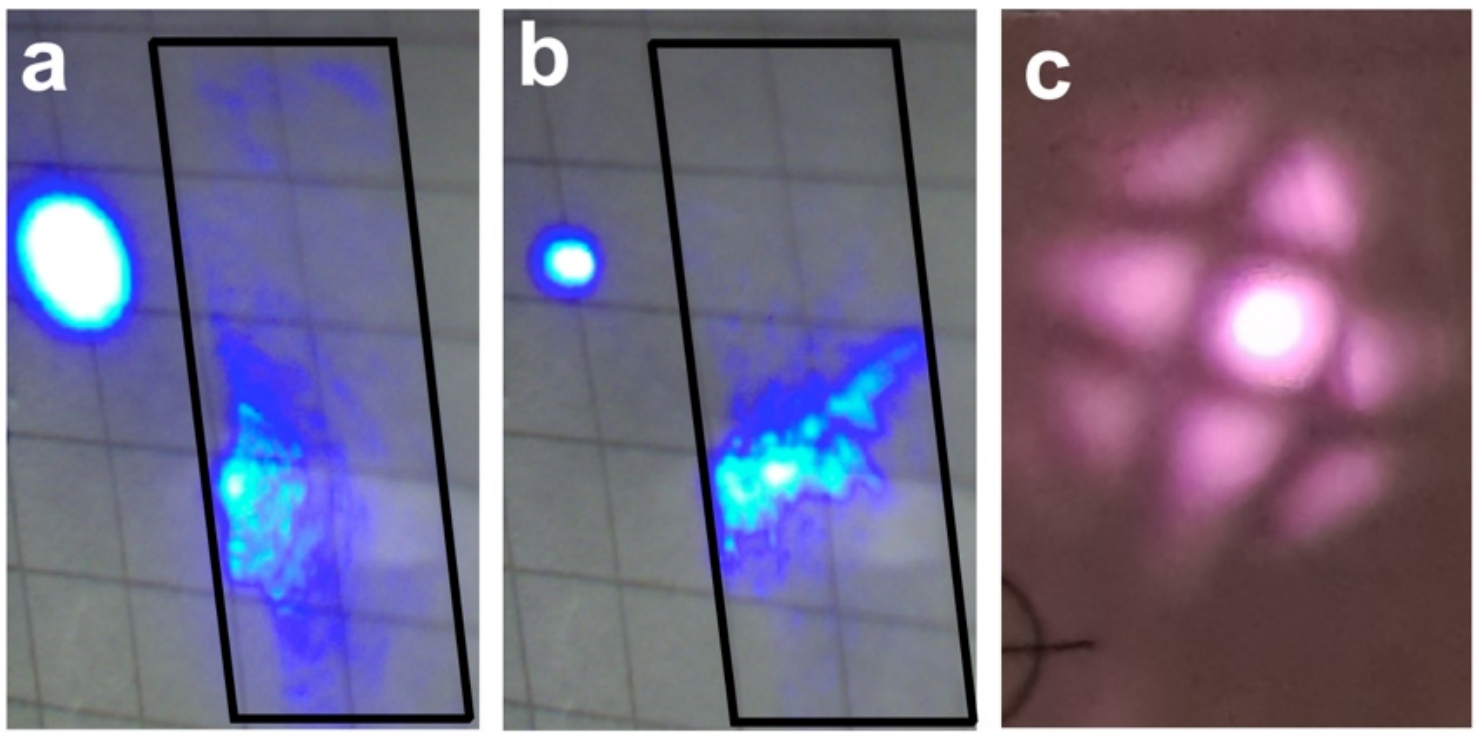

**FIG. S1. (a-b)** Photos of UV-induced fluorescence on a piece of paper (a) with and (b) without a UV lens positioned before the input CF40 UVFS window measured with a camera. The outcoupled UV beam using the in-vacuum rectangular mirror (see tilted black rectangle) is shown with the back-reflection of the UV beam reflected by the input CF40 UVFS window (see circular beam profile). **(c)** Outcoupled 800-nm pump pulse that passed through a meshgrid sample (300 lines/inch) with a 100-μm diameter copper aperture placed in-front of the meshgrid for alignment purposes.

### B. HiRep-UED instrument

The footprint of the HiRep-UED instrument is approximately 1.5-m × 1.0-m. The electron accelerator chamber with the six-way laser in-coupler unit (Doctor X Works BV, formerly AccTec BV; see Fig. 2c) is mounted onto an optical table (3-m × 2-m), and is completely surrounded by X-ray radiation shielding comprised of ≥4-mm thick lead shielding (attenuating the X-ray dosage by four-to-five orders of magnitude reduces the radiation levels outside the shielding to <<1-μSv/hr). The rest of the instrument is supported by a frame (~1.0-m × 0.5-m). This frame is equipped with rubber feet to minimize vibrations from the surrounding area. Each evacuated chamber is mounted on sliding plates attached to two longitudinal profiles that are secured to the frame. This design enables the chambers to be positioned anywhere along the frame, providing flexibility in placement, such as adjusting the distance between the detector and the photocathode. Bellows are incorporated between chambers to provide flexibility in adjusting their positions relative to each other. Each high-vacuum chamber was evacuated by turbomolecular pumps with a base background pressure of $<7 \times 10^{-7}$ mbar in the electron accelerator chamber (Edwards EXT 75DX) and the laser in-coupling six-way cross (Edwards EXT 75DX), and $<3 \times 10^{-8}$ mbar in the radiofrequency (RF) cavity (Edwards nEXT 85D) and sample (Edwards nEXT 300) chambers. An existing 1,600 l/s turbomolecular pump (Edwards STPIXR1606) planned for future gas-phase measurements was capable of achieving $<9 \times 10^{-10}$ mbar base pressure in the sample chamber. All turbomolecular pumps are pre-pumped by a 500 l/s dry vacuum pump (Edwards EPX 500LE) that can reach a base pressure of $<10^{-4}$ mbar. A high vacuum gate valve (MAC N-7557-019) is employed after the electron accelerator to prevent the vacuum from being compromised (and to avoid high voltage breakdown issues) in the accelerator when, for example, operating the quick access door (Kurt J. Lesker) in the sample chamber to change samples. Two out-of-vacuum charge-coupled device (CCD) cameras are used to view the sample holder and to monitor the out-coupled optical pump pulse, both crucial to achieving optimal pump-probe spatial overlap.

Breakdown and arcing issues with this electron accelerator were circumvented through the following steps. The aluminium housing that supports the copper cathode was polished to mirror finish using diamond paste (from 3-μm to 0.25-μm). The high-vacuum chamber supporting the cathode assembly was also electropolished. Replacement of the copper photocathode (Doctor X Works BV) is recommended for every six-to-twelve months of operation, which has a significant impact on the electron beam's emittance properties.[1] After replacing the photocathode or breaking high-vacuum conditions in the accelerator, the photocathode was trained, over a 48-hour period, to hold a voltage of 100-keV (the maximum voltage of the power supply), where many breakdown events occurred to reach 100-keV. Crucially, the conditioning of the electron accelerator was performed with a relatively high current (*i.e.*, 100-μA) and a low voltage rate (*i.e.*, <0.5-kV/minute). Daily stable operation of the electron accelerator at 95-keV was possible by slowly increasing the voltage to 95-keV in steps of 1-kV/minute over ~1.5 hours, held at a current of 10-μA.

### C. General particle tracer simulation of UV pulse duration

Figure S2 shows the effect of varied UV pulse duration on the simulated electron pulse duration (FWHM). Longer UV pulses lead to longer electron pulses for <100 electrons/pulse but have the added advantage that space-charge repulsion is less severe for >>100 electrons/pulse. While using a UV pulse of 90-fs duration leads to longer duration pulses at >$10^4$ electrons/pulse. Ultimately, no significant difference (*i.e.*, order-of-magnitude) effect is seen in the electron pulse duration when comparing the three durations of the UV pulse. The 6D phase-space distribution must be carefully investigated for a more accurate understanding of the subtle differences across the different three UV pulse durations.

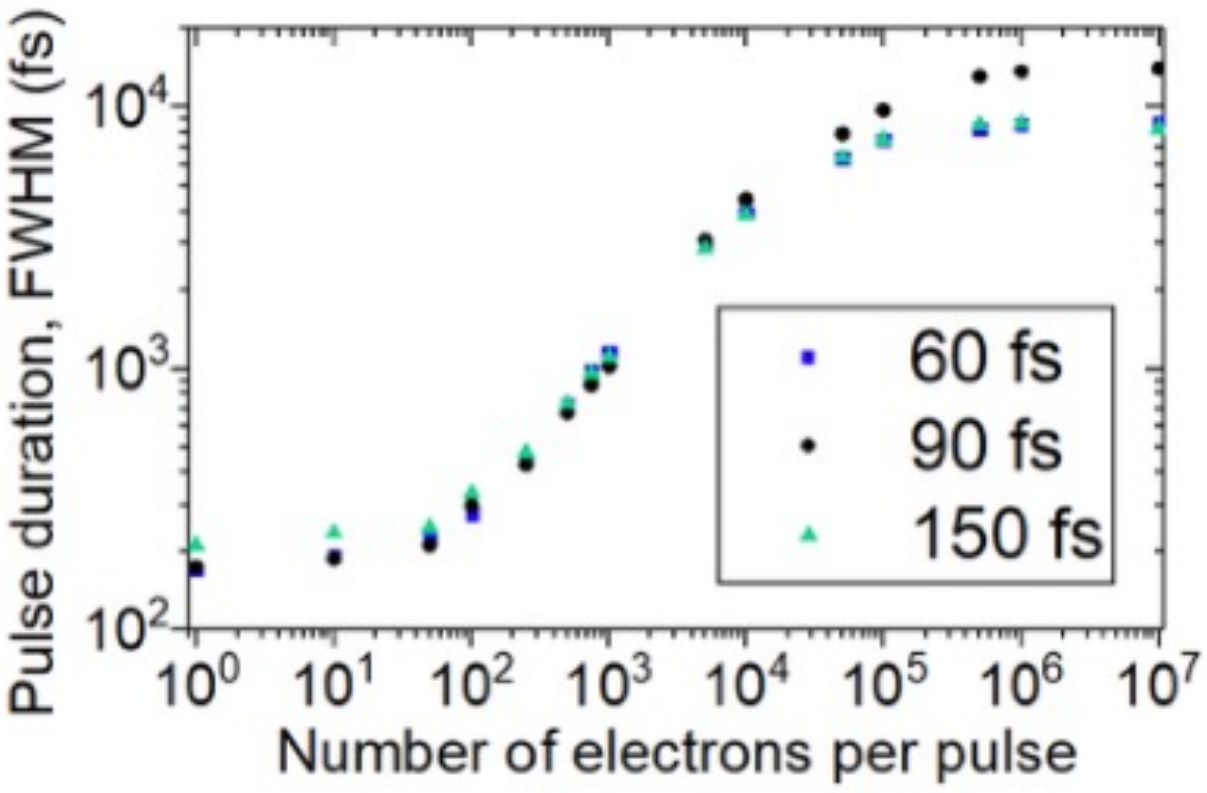

**FIG. S2.** General particle tracer simulations of the temporally uncompressed electron pulse duration (FWHM) for a varied number of electrons in the pulse at the sample generated by an ultraviolet (UV) pulse duration of 60 fs, 90 fs, and 150 fs.

### D. Transverse emittance, coherence and beam diameter of electron beam

The transverse electron beam FWHM diameter was measured at the sample using knife edge scans at various ML1 collimator. The RMS transverse emittance and coherence lengths were then extracted from this data, and shown in Figure S3. The minimum transverse beam diameter is found to be 83 μm FWHM at 9.4 A. This corresponds to an RMS transverse emittance of 3.8 nm·rad and an RMS transverse coherence length of 11 nm.

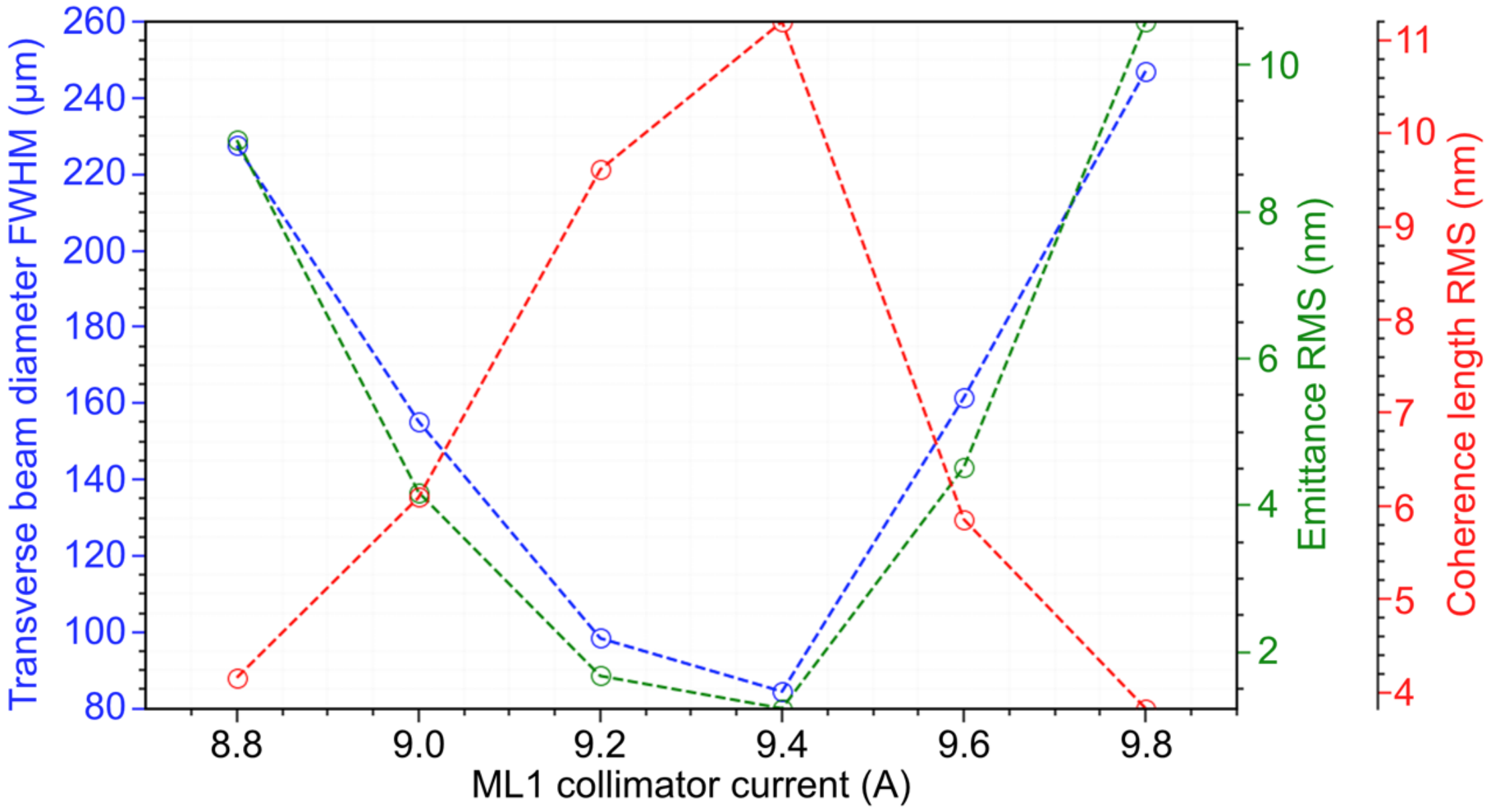

**FIG. S3.** Measured FWHM transverse beam diameter of electron beam (blue) as a function of ML1 collimator current. Knife edge scans were employed at the sample position. The RMS transverse (green) and coherence length (red) were extracted from measured beam diameter. The electron pulse contained 130 electrons and the current of ML2 was fixed to ~1.6 A.

The RMS transverse emittance was determined as follows. Since we are considering the photo-generated bunch to be elliptical and uniformly charged,[2] the normalised thermal emittance[3] is

$$\varepsilon_{\mathrm{n,th}} = \sigma_{\mathrm{x}}\sqrt{\frac{\hbar\omega - \phi_{\mathrm{eff}}}{3mc^2}},$$

where $\sigma_{\mathrm{x}}$ is the RMS transverse beam diameter, $\hbar\omega$ is the photon energy of the optical pulse generating the photoelectrons, $\phi_{\mathrm{eff}}$ is the effective work function of the photocathode.

The linear electric and magnetic fields can focus the bunch after it has experienced a linear velocity chirp given by the space-charge field.[4] This linear position-momentum correlation dominates the transverse properties of the bunch, and the normalized emittance at the beam-waist is therefore

$$\varepsilon_{\mathrm{n,x}} = \frac{1}{mc}\sigma_{\mathrm{i}}\sigma_{p_{\mathrm{x}}}.$$

Furthermore, due to the linear corelation, it is possible to estimate the emittance from a local area of the bunch,[5] following a Gaussian truncation, as

$$\varepsilon_{\mathrm{n,x}} = \frac{1}{mc}\sigma_{\mathrm{x}}\left[\sigma_{p_{\mathrm{x}}}\right]_{\mathrm{loc}},$$

where the local momentum can be determined via the distribution at the beam waist[6]

$$\left[\sigma_{p_{\mathrm{x}}}\right]_{\mathrm{loc}} = \frac{\gamma m v_{\mathrm{z}}}{l_{\mathrm{cam}}}\sigma_{\mathrm{x}},$$

where $\gamma m v_{\mathrm{z}}$ is the momentum longitudinal momentum $p_{\mathrm{z}}$, $l_{\mathrm{cam}}$ is the camera length, with the factor $\sigma_{\mathrm{x}}/l_{\mathrm{cam}}$ considering the beam divergence. From Busch's theorem, the change in transverse momentum, and thus transverse emittance,[7] due to the solenoids is found to be linear with the magnetic field strength, and is given by

$$\varepsilon_{\mathrm{n,x}} = \frac{q_{\mathrm{e}} c_{\mathrm{x}}^{2}}{8mc^{2}}\boldsymbol{B}_{0},$$

noting that the effects of field tails can be neglected as the amplitude of the field decreases outside the solenoid area.